\documentclass[journal]{IEEEtran}
\usepackage{amsfonts,bm,amsmath,comment,color,cite,amssymb,subfigure,multicol,multirow,array}
\usepackage[]{graphicx}
\usepackage[linesnumbered,ruled]{algorithm2e}

\def\bzero{{\mathbf{0}}}
\def\bone{{\mathbf{1}}}
\def\c1{{\textcircled{a}}}
\def\ba{{\mathbf{a}}}
\def\bb{{\mathbf{b}}}
\def\bc{{\mathbf{c}}}
\def\bd{{\mathbf{d}}}
\def\be{{\mathbf{e}}}

\def\bm{{\mathbf{m}}}

\def\bs{{\mathbf{s}}}

\def\bu{{\mathbf{u}}}
\def\bv{{\mathbf{v}}}
\def\bw{{\mathbf{w}}}
\def\bx{{\mathbf{x}}}
\def\by{{\mathbf{y}}}
\def\bz{{\mathbf{z}}}
\def\bA{{\mathbf{A}}}
\def\bB{{\mathbf{B}}}
\def\bC{{\mathbf{C}}}

\def\bI{{\mathbf{I}}}
\def\bJ{{\mathbf{J}}}
\def\bK{{\mathbf{K}}}

\def\bM{{\mathbf{M}}}

\def\bQ{{\mathbf{Q}}}
\def\bR{{\mathbf{R}}}
\def\bS{{\mathbf{S}}}

\def\bU{{\mathbf{U}}}
\def\bV{{\mathbf{V}}}
\def\bW{{\mathbf{W}}}
\def\bX{{\mathbf{X}}}
\def\bY{{\mathbf{Y}}}

\def\complexC{{\mathbb{C}}}

\def\bzero{{\mathbf{0}}}
\def\bone{{\mathbf{1}}}

\def\tr{{\textrm{tr}}}

\def\NT{{N_\textrm{T}}} 
\def\NR{{N_\textrm{R}}} 
\def\txT{{\textrm{T}}} 
\def\txR{{\textrm{R}}} 
\def\txJ{{\textrm{J}}} 
\def\txn{{\textrm{n}}} 
\def\txt{{\textrm{t}}} 
\def\txc{{\textrm{c}}} 
\def\txu{{\textrm{u}}} 

\renewcommand\Re{{\textrm{Re}}}
\renewcommand\vec{{\textrm{vec}}}

\newcommand{\eg}{{\it e.g.}}
\newcommand{\ie}{{\it i.e.}}

\begin{document}
%
\title{Polyphase Waveform Design for MIMO Radar Space Time Adaptive Processing}
\author{Bo~Tang,~
             Jonathan~Tuck,~\IEEEmembership{Member,~IEEE,}
             and Peter~Stoica,~\IEEEmembership{Fellow,~IEEE}
\thanks{Bo Tang is with the College of Electronic Engineering, National University of Defense Technology, Hefei 230037, China
        (email: tangbo06@gmail.com). }
\thanks{Jonathan~Tuck is with the Department of Electrical Engineering, Stanford University, Stanford, CA 94305, USA (email: jonathantuck@stanford.edu).}
\thanks{Peter Stoica is with the Department of Information Technology, Uppsala University, Uppsala, SE 75105, Sweden (email: ps@it.uu.se).}
}

\markboth{Preprint}%
{Shell \MakeLowercase{\textit{et al.}}: Bare Demo of IEEEtran.cls for Journals}

\maketitle

\begin{abstract}
We consider the design of polyphase waveforms for ground moving target detection with airborne multiple-input-multiple-output (MIMO) radar. Due to the constant-modulus and finite-alphabet constraint on the waveforms, the associated design problem is non-convex and in general NP-hard. To tackle this problem, we develop an efficient algorithm based on relaxation and cyclic optimization. Moreover, we exploit a reparameterization trick to avoid the significant computational burden and memory requirement brought about by relaxation. We prove that the objective values during the iterations are guaranteed to converge. Finally, we provide an effective randomization approach to obtain polyphase  waveforms from the relaxed solution at convergence.
Numerical examples show the effectiveness of the proposed algorithm for designing polyphase  waveforms. 
\end{abstract}


\section{Introduction}
Multiple-input-multiple-output (MIMO) radar refers to a system with multiple transmitters and multiple receivers. Compared with the traditional phased-array radar, MIMO radar can transmit multiple independent probing waveforms simultaneously, enabling an increased waveform diversity. According to the array configurations, MIMO radar can be classified into two types: statistical MIMO radar \cite{Li2009mimoBook,Haimovich2008statMIMOintro} and coherent MIMO radar \cite{Li2009mimoBook,Li2007mimoIntroduction}. Statistical MIMO radar, which is also called distributed MIMO radar, refers to a system with widely separated transmit/receive antennas. The spatial diversity provided by statistical MIMO radar can be used to overcome the target amplitude  fluctuation \cite{Haimovich2008statMIMOintro,Fishler2006mimo} and attain a higher localization accuracy \cite{Haimovich2008statMIMOintro,Godrich2010Localization}.  Different from statistical MIMO radar, the antennas of coherent MIMO radar are colocated. Thus, coherent MIMO radar is also known as colocated MIMO radar. Compared with a phased array radar, coherent MIMO radar can achieve a larger number of degrees of freedom (DOF) \cite{Bliss2003mimoDOF} and improved parameter identifiability \cite{Li2007mimoiden}.

Owing to its superior target detection and parameter estimation performance, MIMO radar has found many recent applications including autonomous driving \cite{bilik2016automotive},  remote sensing \cite{ender2009mimoSAR}, landmine detection \cite{lim2015FLGPR}, and disaster management \cite{klare2014disastermanagement}, just to name a few. In this paper, we focus on the use of an airborne colocated MIMO radar for ground moving target detection. In this down-looking scenario, the strong ground clutter can obscure the weak target returns. In addition, the radar platform motion makes the clutter widely spread not only in range and azimuth, but also in Doppler, resulting in the difficulty of detecting slowly moving targets (\ie, the target Doppler frequencies are close to that of clutter). To improve the target detection performance of an airborne radar in the presence of clutter, space time adaptive processing (STAP) with a single-input-multiple-output (SIMO) array (\ie, phased array) has been proposed \cite{Brennan1973stapPaper,Guerci2003stapBook,Ward1995stapReport}.
The key of STAP is to use an adaptive multi-dimensional filter to form a deep notch on the clutter ridge and thus suppress the clutter effectively.
In recent years, STAP has been extended to MIMO radar systems \cite{Mecca2006mimoSTAP,ChenVaidyanathan2008MIMOSTAP,ForsytheBliss2010mimoGMTI,XueLiStoica2010mimoGMTI}. Compared with conventional SIMO STAP, MIMO STAP can attain a better minimum-detectable-velocity (MDV) performance; that is, such a radar system is more able to detect slowly moving targets in the presence of clutter.

In this paper, we consider STAP for airborne MIMO radar and, in particular, designing probing waveforms for the MIMO radar. We assume that we have \emph{a priori } knowledge about the interference. This assumption is justified for adaptive radar systems, which can get such prior knowledge from other sensors or databases (\eg, synthetic aperture radar (SAR), land cover land use (LCLU)
data, digital terrain elevation data (DTED), etc.), or by revisiting the adaptive estimation results of previous scans. We use the signal-to-interference-plus-noise-ratio (SINR)  as the design metric. Note that a related waveform design problem based on SINR maximization for MIMO radar in the presence of clutter has been discussed in \cite{Chen2009WaveformDesign,Tang2014MIMOdesign,CuiLi2014MIMOsim,Karbasi2015MIMOdesign,my2016JointDesign,my2016wcm,Cui2017SpaceTime} (see also \cite{Aubry2013simClutter,Aubry2012Cognitivedesign,Naghsh2014robustdesign,Stoica2012Optimization,Soltanalian2013Joint} and the references therein for similar waveform design problems for single-input-single-output (SISO) radar). In the cited references, algorithms have been developed for designing waveforms under certain  practical constraints, including energy constraint, constant-modulus constraint, and a similarity constraint. However, the phases of the MIMO radar waveforms provided by these algorithms are continuous (over $[0, 2\pi)$ or a smaller interval), which makes them more difficult to implement in real radar systems. Indeed, the design of polyphase waveforms, whose phases belong to  a finite-alphabet set, is highly desirable in practical radar systems.

To tackle the polyphase waveform design problem for MIMO STAP, we develop  an efficient algorithm based on relaxation and cyclic optimization.
Through the use of a proper reparameterization idea,  the proposed algorithm avoids the significant computational burden and memory requirement of semidefinite programming (SDP). In addition, by using Dinkelbach's transform and the minorization-maximization (MM) technique, the problems that need to be solved admit a closed-form solution at each inner iteration. Moreover, we propose an improved randomization approach to design high-quality polyphase waveforms from the relaxed solution provided by the proposed algorithm.

The rest of this paper is organized as follows. Section \ref{sec:SigModel} establishes the signal model and formulates the waveform design problem.
Section \ref{sec:AlgorithmDesign} develops an algorithm for designing polyphase waveforms for airborne MIMO radar in the presence of clutter.
Section \ref{sec:Discussions} analyzes the convergence and computational complexity of the proposed algorithm.
Section \ref{sec:Examples} provides numerical examples to demonstrate the performance of the proposed algorithm.
Finally, we conclude the paper in Section \ref{sec:Conclusion}.

\emph{Notation}:
Throughout this paper, matrices are denoted by bold uppercase letters and vectors are denoted by bold lowercase letters.
$\mathbb{C}^{m\times n}$ and $\mathbb{C}^{k}$ are the sets of ${m\times n}$ matrices and $k\times1$ vectors with complex-valued entries, respectively.
$\bI_M$ denotes the $M\times M$ identity matrix. $\bone_{M\times N}$ and $\bzero_{M\times N}$ denote the $M\times N$ matrices of ones and zeros, respectively.
Superscripts $(\cdot)^{T}$ and $(\cdot)^{H}$ denote transpose and conjugate transpose, respectively.
$\textrm{tr}(\cdot)$ denotes the trace of a matrix.
$\otimes$ represents the Kronecker  product.
$\|\bx\|_2$ denotes the Euclidian norm of the vector $\bx$.
${\vec}(\bX)$ indicates the vector obtained by column-wise stacking of the entries of $\bX$.
$\textrm{diag}(\bX)$ denotes the vector formed by the diagonal elements of $\bX$.
$\Re(\bX)$ denotes the real part of the matrix $\bX$ (element-wise).
$\arg(x)$ represents the argument of $x$.
$\textrm{BlkDiag}([\bA;\bB])$ denotes the block-diagonal matrix formed by the matrices $\bA$ and $\bB$.
The notation $\bA \succ \bB $ ($\bA \succeq \bB$) means that $ \bA - \bB$ is positive definite (semi-definite).
$\bx\sim \mathcal{CN}(\bm,\bR)$ means that $\bx$ obeys a circularly symmetric complex Gaussian distribution with mean $\bm$ and covariance matrix $\bR$.
$\mathbb{E}(x)$ denotes the expectation of the random variable $x$.
Finally, $\lfloor x\rceil$ and $\lfloor x\rfloor$ return the nearest integer to $x$ and the nearest integer less than or equal to $x$, respectively.

\newtheorem{Def}{Definition}
\newtheorem{lemma}{Lemma}
\newtheorem{theorem}{Theorem}
\newtheorem{Prop}{Proposition}

\section{Signal Model and Problem Formulation} \label{sec:SigModel}
Consider an airborne  colocated MIMO radar system with $N_\txT$ transmit antennas and $N_\txR$ receive antennas.
Let $\bs_n \in \mathbb{C}^{L}$ denote the (discrete-time) waveform of the $n$th transmitter and let $\bS = [\bs_1,\bs_2, \ldots, \bs_{N_\txT}]^T\in \mathbb{C}^{\NT \times L}$ denote the transmit waveform matrix, where $L$ is the code length. Assume that the radar system transmits a burst of $M$ pulses in a coherent processing interval (CPI) with a pulse repetition frequency (PRF) denoted by $f_r$. In addition, the transmitted waveforms are repeated from pulse to pulse. For a down-looking airborne radar, the received signal consists of a possible target return, clutter (due to reflections from ground or sea), and possible jamming signals. Before formulating the waveform design problem, we establish the signal model similarly to what has been done in \cite{my2016JointDesign}.
\subsection{Target}
Assume that the transmit waveforms are narrowband.  Under the far-field assumption, the target return at the receive array from the $m$th pulse ($m=1, \ldots, M$)  can be written as
\begin{equation} \label{Eq:targetModel}
  \bY_{\textrm{t},m} = \alpha_\textrm{t}e^{j(m-1)\omega_\textrm{t}}\ba_\txR(\theta_\textrm{t})\ba_\txT^T(\theta_\textrm{t})\bS,
\end{equation}
where $\alpha_\textrm{t}$ is the target amplitude, $\omega_\textrm{t }= 2\pi f_\textrm{t}$, $f_\textrm{t}$ is the normalized target Doppler frequency, $\theta_\textrm{t}$ is the target direction of arrival (DOA),  $\ba_\txT(\theta_\textrm{t}) \in \mathbb{C}^{\NT}$ and $\ba_\txR(\theta_\textrm{t}) \in \mathbb{C}^{\NR}$ are the transmit array steering vector and the receive array steering vector at $\theta_\textrm{t}$, respectively.
(For a detailed derivation of \eqref{Eq:targetModel}, we refer to \cite{my2016JointDesign}.)

If the transmit and receive arrays are both uniform linear arrays (ULA), $\ba_\txT(\theta_\textrm{t})$ and  $\ba_\txR(\theta_\textrm{t})$ can be written as
\begin{align*}
\ba_\txT(\theta_\textrm{t}) &= [1, \ldots, e^{j2\pi (\NT-1)d_\txT \sin(\theta_\textrm{t})/\lambda }]^T,\\
  \ba_\txR(\theta_\textrm{t}) &= [1, \ldots, e^{j2\pi (\NR-1)d_\txR \sin(\theta_\textrm{t})/\lambda }]^T,
\end{align*}
where $d_\txT$ and $d_\txR$ are the inter-element spacing of the transmitters and the receivers, respectively,  and $\lambda$ is the wavelength.

Let $\by_{\textrm{t},m} = \vec({\bY_{\textrm{t},m}})$. Then
\begin{equation}
  \by_{\textrm{t},m} =  \alpha_\textrm{t }e^{j(m-1)\omega_\textrm{t}}(\bI_L \otimes\bA(\theta_\textrm{t}) )\bs,
\end{equation}
where $\bA(\theta_\textrm{t}) = \ba_\txR(\theta_\textrm{t})\ba_\txT^T(\theta_\textrm{t})$, and $\bs = \vec(\bS)$.
Define $\by_{\textrm{t}} = [\by_{\textrm{t},1}^{T},\ldots, \by_{\textrm{t},M}^T]^T \in \mathbb{C}^{LM\NR}$, which denotes the target return associated with the $M$ pulses in a CPI. Then $\by_\textrm{t}$ can be written as
\begin{equation}
  \by_\textrm{t} = \alpha_\textrm{t }(\bd(\omega_\textrm{t})\otimes\bI_L \otimes\bA(\theta_\textrm{t} ))\bs,
\end{equation}
where $\bd(\omega_\textrm{t}) = [1, \ldots, e^{j(M-1)\omega_\textrm{t}}]^T$, which is the temporal steering vector at the Doppler frequency $f_\textrm{t}$.

\subsection{Clutter}
In general, clutter refers to unwanted reflections (\ie, an interference that is dependent on the transmit waveforms), which can be much stronger than the target return and thus harmful to target detection.
To facilitate the waveform design, we consider the $2P+1$ clutter rings in the vicinity of the cell under test with each ring being at the same range to radar. For each clutter ring, we approximate the clutter by a superposition of $N_c$ clutter patches, which are evenly distributed in azimuth (note that such an approximation is widely used in STAP, see, \eg, \cite{Ward1995stapReport,Guerci2003stapBook}):
\begin{equation}
  \by_{\textrm{c}} = \sum_{p = -P}^{P}\sum_{k=1}^{N_\textrm{c}} \alpha_{\textrm{c},p,k}(\bd(\omega_{\textrm{c},p,k})\otimes\bJ_p^T \otimes\bA(\theta_{\textrm{c},p,k}))\bs,
\end{equation}
where $\alpha_{\textrm{c},p,k}$, $\omega_{\textrm{c},p,k}$, and $\theta_{\textrm{c},p,k}$ are the amplitude, the Doppler frequency (in radians), the DOA of the $k$th clutter patch in the $p$th clutter ring, respectively (for $p=0$, the clutter has the same range as the target),  and $\bJ_p = \bJ_{-p}^T \in \mathbb{C}^{L\times L}$ is the shift matrix defined by 
\begin{equation*}
  \bJ_{p}(i,j) =
  \begin{cases}
    1, \textrm{if} \ i-j+p = 0\\
    0, \textrm{if} \ i-j+p \neq 0
  \end{cases}.
\end{equation*}
Note that the Doppler frequencies of the clutter patches are coupled with their DOAs, \eg, for an airborne radar with the platform velocity vector perfectly aligned with the antenna array axis, 
\begin{equation*}
  \omega_{\textrm{c},p,k} = \frac{4\pi v_a}{f_r\lambda} \sin(\theta_{\textrm{c},p,k}),
\end{equation*}
where $v_a$ denotes the platform velocity.
\subsection{Jamming}
Jamming refers to a signal-independent interference caused by  hostile electronic countermeasures. The jamming signal can be modeled as
\begin{equation}
  \by_\textrm{J} = [\by_{\textrm{J},1}^T, \ldots, \by_{\textrm{J},M}^T]^T,
\end{equation}
where $\by_{\textrm{J},m}$ denotes the received jamming signal for the $m$th pulse, which can be written as
\begin{equation}
  \by_{\textrm{J},m} = \sum_{j=1}^{N_\txJ}\ba_\txR(\theta_{\txJ,j})\otimes \boldsymbol{\epsilon}_{j,m},
\end{equation}
with $N_\txJ$ denoting the number of present jammers, $\theta_{\txJ,j}$ and $\boldsymbol{\epsilon}_{j,m} \in \mathbb{C}^{L}$ representing the DOA and the jamming waveform of the $j$th jammer, respectively.

\subsection{Problem Formulation}
Detecting the target is tantamount to  the following hypothesis testing problem:
\begin{equation}
  \begin{cases}
    \mathcal{H}_0:& \by = \by_\txu\\
    \mathcal{H}_1:& \by = \by_\txt + \by_\txu %
\end{cases},
\end{equation}
where $\by_\txu = \by_\txc + \by_\txJ + \by_\txn$ denotes the undesired components, and $\by_\txn$ denotes the noise in the receiver. It is a well established fact in STAP that to enhance the target detection performance, we can design a multi-dimensional filter, denoted by $\bw \in \mathbb{C}^{LM\NR}$, to maximize the SINR. Using such a multi-dimensional filter leads to the following expression for the SINR
\begin{align} \label{eq:SINR}
  \textrm{SINR}
  &= \frac{|\bw^H \by_t|^2}{\bw^H \mathbb{E}(\by_\txu\by_\txu^H) \bw } \nonumber\\
  &= \frac{|\alpha_\txt|^2|\bw^H \bv_\txt(\bs)|^2}{\bw^H \bR_\txu(\bs) \bw } ,
\end{align}
where $\bv_\txt(\bs) = (\bd(\omega_\textrm{t})\otimes\bI_L \otimes\bA(\theta_\textrm{t} ))\bs$, and $\bR_\txu(\bs) = \mathbb{E}(\by_\txu\by_\txu^H)$ denotes the interference-plus-noise covariance matrix. Assume that the clutter, the jamming signal, and the receiver noise are uncorrelated with each other. Then  $\bR_\txu$ can be written as
\begin{equation}
  \bR_\txu(\bs) = \bR_\txc(\bs) + \bR_\txJ + \bR_\txn,
\end{equation}
where
\begin{itemize}
  \item $\bR_\txc(\bs) = \mathbb{E}(\by_\txc \by_\txc ^H)$ denotes the clutter covariance matrix. In this paper, we assume that the reflection from different clutter patches are uncorrelated. Under this assumption, the clutter covariance matrix is given by
      \begin{equation}
        \bR_\txc(\bs) = \sum_{p = -P}^{P}\sum_{k=1}^{N_\textrm{c}} \sigma^2_{\textrm{c},p,k}\bv_{\txc,p,k}(\bs) \bv^H_{\txc,p,k}(\bs),
      \end{equation}
      where $\sigma^2_{\textrm{c},p,k} = \mathbb{E}(|\alpha_{\textrm{c},p,k}|^2)$ denotes the average power of the $k$th clutter patch in the $p$th range ring, and $\bv_{\txc,p,k}(\bs) = (\bd(\omega_{\textrm{c},p,k})\otimes\bJ_p^T \otimes\bA(\theta_{\textrm{c},p,k}))\bs$.
  \item $\bR_\txJ = \mathbb{E}(\by_\txJ \by_\txJ^H)$ denotes the jammer covariance matrix. If we assume barrage noise jamming, and that the received jamming samples are uncorrelated in both fast time and slow time, $\bR_\txJ $ can be written as
      \begin{equation}
        \bR_\txJ = \left(\sum_{j=1}^{N_\txJ} \sigma_{\txJ,j}^2 \ba_\txR(\theta_{\txJ,j}) \ba^H_\txR(\theta_{\txJ,j})\right)\otimes \bI_{LM},
      \end{equation}
      where $\sigma_{\txJ,j}^2$ denotes the average power of the $j$th jammer.
  \item $\bR_\txn = \mathbb{E}(\by_\txn\by_\txn^H)$ denotes the noise covariance matrix. For white noise with power level $\sigma^2$, $\bR_\txn= \sigma^2\bI_{LM\NR}$.
\end{itemize}

It can be verified that the optimal filter maximizing the SINR is given by
\begin{equation}\label{eq:OptimalFilter}
  \bw_{\textrm{opt}} = \gamma \bR_\txu^{-1}(\bs) \bv_\txt(\bs),
\end{equation}
where $\gamma$ is an arbitrary nonzero constant (\eg, if we require that $\bw^H \bv_\txt = 1$, then $\gamma = (\bv_\txt^H \bR_\txu^{-1}\bv_\txt)^{-1}$ and the filter in \eqref{eq:OptimalFilter} is the so-called minimum variance distortionless response (MVDR) beamformer).
Using \eqref{eq:OptimalFilter}, the SINR in  \eqref{eq:SINR} becomes
\begin{equation}
  \textrm{SINR} = |\alpha_\txt|^2 (\bv_\txt(\bs))^H \bR_\txu^{-1}(\bs)\bv_\txt(\bs).
\end{equation}
Note the dependency of the SINR on the transmit waveforms $\bs$  (through both $\bv_\txt$ and $\bR_\txu$). We aim to design $\bs$
to maximize the SINR and improve the detection performance of MIMO radar, especially for the slowly moving targets.

In practical radar systems, the available transmit energy is limited. Thus we impose the constraint $\bs^H \bs = e_t$. In addition, to allow the radio frequency amplifier to operate at maximum efficiency and avoid unnecessary nonlinear effects in transmitters, constant-modulus waveforms are of great interest in practice. Under the constant-modulus constraint, we have $|s_n(l)| = \sqrt{p_s}, n=1,\ldots,\NT, l=1, \ldots, L$, where $p_s = e_t/(L\NT)$. Since such waveforms are determined by their phases, we also call them phase-coded waveforms. Moreover, the phases of radar waveforms are usually restricted to lie in a finite-alphabet set, in which case they are called polyphase waveforms. For polyphase waveforms, their phases, denoted by $\{\{\phi_n(l)\}_{n=1}^{\NT}\}_{l=1}^{L}$, belong to the set $\mathcal{S} = \{0, \Delta\phi ,\ldots, (D-1)\Delta\phi \}$,
where $D\geq 2$ is the number of phases in the set, and $\Delta\phi = 2\pi/D$.
If $D=2$, $\mathcal{S}  = \{0, \pi\}$, and the transmit waveforms ($\bs_1,\bs_2,\ldots, \bs_{\NT}$) are called binary or 1-bit waveforms, among which the Barker code \cite{Levanon2004RadarSignals} is one of the most well-known.

Thus, to design polyphase waveforms that maximize the SINR, we formulate the following optimization problem:
\begin{align}\label{eq:Problem}
  \max_{\bs}& \ (\bv_\txt(\bs))^H \bR_\txu^{-1}(\bs)\bv_\txt(\bs) \nonumber \\
  \textrm{s.t.}& \ |s_n(l)| = \sqrt{p_s},  \phi_n(l) = \arg(s_n(l)) \in \mathcal{S}, \nonumber \\
                    &\  n=1,\ldots,\NT, l=1, \ldots, L.
\end{align}

Note that we have assumed that prior knowledge on the interference is available in \eqref{eq:Problem}. This is justified in adaptive radar systems. In such systems, we can estimate $\bR_{\txJ\txn}$ by operating the radar system in a passive mode.  Regarding the clutter, prior knowledge about it can be provided by the platform motion and the array geometry, or by the estimation results from previous scans (\eg, the clutter profile can be accurately estimated by the iterative adaptive approach proposed in \cite{Li2010IAA}).


\section{Algorithm Derivation} \label{sec:AlgorithmDesign}
It is apparent that the optimization problem in \eqref{eq:Problem} is non-convex, due to the constraints on the waveforms.
To proceed, we return to \eqref{eq:SINR} and formulate the optimization problem in \eqref{eq:Problem} as a joint design problem of the transmit waveform $\bs$  and the receive filter $\bw$:
\begin{align}\label{eq:JointDesign}
  \max_{\bw,\bs}& \  \frac{|\bw^H \bv_\txt(\bs)|^2}{\bw^H \bR_\txu(\bs) \bw } \nonumber \\
  \textrm{s.t.}& \ |s_n(l)| = \sqrt{p_s},  \phi_n(l) \in \mathcal{S}, \nonumber \\
                    &\  n=1,\ldots,\NT, l=1, \ldots, L.
\end{align}
The optimal $\bw$  of course is given by \eqref{eq:OptimalFilter} and if we substitute it into \eqref{eq:JointDesign}, the optimization problem \eqref{eq:JointDesign} reduces to  \eqref{eq:Problem}. However, as we will show later on, the formulation in \eqref{eq:JointDesign} is more suitable for  developing an efficient algorithm.

Let $\bV_\txt = \bd(\omega_\textrm{t})\otimes\bI_L \otimes\bA(\theta_\textrm{t} )$ and $\bV_{\txc,k,p} = \bd(\omega_{\textrm{c},p,k})\otimes\bJ_p^T \otimes\bA(\theta_{\textrm{c},p,k})$. Then
\begin{equation}\label{eq:CostDeno}
  |\bw^H \bv_\txt(\bs)|^2 = \bw^H \bV_\txt  \bs\bs^H \bV_\txt^H \bw
\end{equation}
and
\begin{equation}\label{eq:CostNume}
  \bR_\txc(\bs) = \sum_{p = -P}^{P}\sum_{k=1}^{N_\textrm{c}} \sigma^2_{\textrm{c},p,k} \bV_{\txc,k,p} \bs \bs^H \bV_{\txc,k,p}^H.
\end{equation}

Define $\bR_s = \bs\bs^H$. It is easy to check that $\bR_s \succeq \bzero$ and $\textrm{diag}(\bR_s) = p_s\cdot \bone_{L\NT\times 1}$. In addition,  $\textrm{rank}(\bR_s)=1$ and under the finite-alphabet constraint, the phases of the elements of $\bR_s$ belong to $\{-(D-1)\Delta\phi ,\ldots, (D-1)\Delta\phi \}$. Moreover, using \eqref{eq:CostDeno} and \eqref{eq:CostNume}, the objective function in  \eqref{eq:JointDesign} can be written as
\begin{equation}
  \frac{\bw^H \bV_\txt \bR_s  \bV_\txt^H \bw}{\bw^H  \bR_\txu(\bR_s) \bw},
\end{equation}
where $\bR_\txu(\bR_s) = \bR_\txc(\bR_s) + \bR_{\txJ\txn}$, $\bR_\txc(\bR_s) = \sum_{p = -P}^{P}\sum_{k=1}^{N_\textrm{c}} \sigma^2_{\textrm{c},p,k} \bV_{\txc,k,p} \bR_s \bV_{\txc,k,p}^H$, and $\bR_{\txJ\txn} = \bR_{\txJ} + \bR_{\txn}$ is the jammer-plus-noise covariance matrix.
Dropping the rank and the element constraints on $\bR_s$, we obtain the following relaxation of the problem \eqref{eq:JointDesign}:
\begin{align}\label{eq:JointDesignRelax}
  \max_{\bw,\bR_s}& \    \frac{\bw^H \bV_\txt \bR_s  \bV_\txt^H \bw}{\bw^H  \bR_\txu(\bR_s) \bw}\nonumber \\
  \textrm{s.t.}& \ \textrm{diag}(\bR_s) = p_s\cdot \bone_{L\NT\times 1}, \bR_s \succeq \bzero.
\end{align}

Similar to \cite{my2016JointDesign} we tackle the optimization problem in \eqref{eq:JointDesignRelax} in a cyclic way: at every cycle,
the solution to \eqref{eq:JointDesignRelax} with respect to $\bw$ is closed form for a fixed $\bR_s$, and vice-versa; the globally optimal $\bR_s$ given $\bw$ can be obtained by solving an equivalent SDP (which can be obtained via Charnes-Cooper transform \cite{charnes1962programming}, see \cite{my2016JointDesign} and \eqref{eq:DesignMs} for the details). Using a relaxed solution provided by this algorithm, we can obtain polyphase waveforms after proper randomization. Since the algorithm involves cyclic optimization and semidefinite relaxation (SDR), we refer it as ``CycSDR" in the sequel.
However, due to the high computational complexity and the significant memory requirement of solvers for the SDP, CycSDR algorithm is time-consuming and becomes prohibitive for designing long waveforms (\eg, $L\NT>500$). Thus, in this paper we derive an efficient algorithm for \eqref{eq:JointDesignRelax}. To this end, we introduce $\bU_s \in \mathbb{C}^{r\times L\NT}$, and we assume that $\bU_s$ satisfies
\begin{equation}\label{eq:repara}
  \bU_s^H \bU_s \approx \bR_s,
\end{equation}
where $r\leq L\NT$ is the largest possible rank of $\bU_s$. (See Subsection \ref{Subsec:LowRank} for a discussion on $r$.)
Using this reparameterization, we can eliminate the positive semidefinite constraint and the first constraint becomes
\begin{equation}
  \|\bu_l\|_2^2 = p_s, \quad l=1,\ldots, L\NT,
\end{equation}
where $\bu_l \in \mathbb{C}^{r}$ is the $l$th column of $\bU_s$.

Therefore, we reformulate the optimization problem in \eqref{eq:JointDesignRelax} as follows:
\begin{align}\label{eq:JointDesignRepara}
  \max_{\bw,\bU_s}& \    \frac{\bw^H \bV_\txt \bU_s^H\bU_s  \bV_\txt^H \bw}{\bw^H \bR_\txu(\bU_s) \bw}\nonumber \\
  \textrm{s.t.}& \ \|\bu_l\|_2^2 = p_s, \quad l=1,\ldots, L\NT,
\end{align}
where
\begin{equation}\label{eq:RuU}
 \bR_\txu(\bU_s) =  \sum_{p = -P}^{P}\sum_{k=1}^{N_\textrm{c}} \sigma^2_{\textrm{c},p,k} \bV_{\txc,k,p} \bU_s^H \bU_s \bV_{\txc,k,p}^H + \bR_{\txJ\txn}.
\end{equation}

In what follows, we propose a cyclic optimization method to tackle the optimization problem \eqref{eq:JointDesignRepara}. Different from the algorithms in \cite{my2016JointDesign}, the proposed algorithm cyclically optimizes $\bw$ and $\bU_s$. Next we provide solutions for each step of the cyclic algorithm. For notational simplicity, we omit the superscripts in the iterations if doing so does not cause any confusion.
\subsection{Optimization of $\bw$ for Fixed $\bU_s$}
Assume $\bU_s$ is fixed. Define $\bQ_{\txt}(\bU_s) = \bV_\txt \bU_s^H\bU_s  \bV_\txt^H$.
Then the optimization problem \eqref{eq:JointDesignRepara} is equivalent to
\begin{align}\label{eq:DesignWeight}
  \max_{\bw}& \    \frac{\bw^H \bQ_{\txt}(\bU_s) \bw}{\bw^H \bR_\txu(\bU_s)\bw}.
\end{align}
The objective function is a generalized Rayleigh quotient \cite{seber2008matrix}, and it is well known that the maximum is attained at
\begin{equation}\label{eq:Optimalw}
  \bw = \mathcal{P}(\bQ_{\txt}(\bU_s),\bR_\txu(\bU_s)),
\end{equation}
where $\mathcal{P}(\bA,\bB)$ denotes the eigenvector associated with the largest generalized eigenvalue of the matrix pair $(\bA, \bB)$.

\subsection{Optimization of $\bU_s$ for Fixed $\bw$}
Assume $\bw$ is fixed. Define $\bQ_{\txt}(\bw) = \bV_\txt^H \bw \bw^H \bV_\txt$
and
\begin{equation} \label{eq:Ruw}
  \bR_\txu(\bw) = \sum_{p = -P}^{P}\sum_{k=1}^{N_\textrm{c}} \sigma^2_{\textrm{c},p,k} \bV_{\txc,k,p}^H\bw \bw^H\bV_{\txc,k,p} + \beta(\bw) \bI_{L\NT},
\end{equation}
where $\beta(\bw) = (\bw^H\bR_{\txJ\txn}\bw)/e_t$.
Then, as shown in Appendix \ref{Apd:A}, the optimization problem \eqref{eq:JointDesignRepara} is equivalent to
\begin{align}\label{eq:DesignU}
  \max_{\bU_s}& \    \frac{\tr(\bU_s \bQ_{\txt}(\bw) \bU_s^H)}{\tr(\bU_s \bR_\txu(\bw)  \bU_s^H)}\nonumber \\
  \textrm{s.t.}& \ \|\bu_l\|_2^2 = p_s, \quad l=1,\ldots, L\NT.
\end{align}

The optimization problem in \eqref{eq:DesignU} belongs to the class of fractional programming problems. We propose an algorithm to tackle the problem in \eqref{eq:DesignU}. The proposed algorithm is based on Dinkelbach's transform \cite{dinkelbach1967fractional}  and uses MM to deal with the quadratic programming problem after the transform (we refer to \cite{Hunter2004MM} for a tutorial introduction to MM). As shown below, we can obtain a closed-form solution at every iteration of the algorithm.

Let $\bU_s^{(n,k)}$ denote the value of $\bU_s$ at the $(n,k)$th  iteration and $x^{(n,k)}$ denote the associated SINR, where we use $n$ to denote the outer loop for cyclic optimization and $k$ to denote the inner loop for tackling the fractional programming problem with Dinkelbach's transform. Using Dinkelbach's transform, we consider the following optimization problem at the $(n,k+1)$th iteration:
\begin{align}\label{eq:GeneralQuadProg}
  \max_{\bU_s}& \ \tr(\bU_s \bK^{(n,k)}\bU_s^H) \nonumber \\
  \textrm{s.t.}& \ \|\bu_l\|_2^2 = p_s, \quad l=1,\ldots, L\NT,
\end{align}
where $\bK^{(n,k)} = \bQ_{\txt}(\bw^{(n+1)}) - x^{(n,k)}\bR_\txu(\bw^{(n+1)})$. This optimization problem is still non-convex due to the norm constraints on the columns of $\bU_s$, and because $\bK^{(n,k)}$ is not necessarily positive semidefinite.
In the following, we use MM to tackle this problem. To this end, we note that
\begin{equation}\label{eq:majorInequ}
  \tr((\bU_s - \bU_s^{(j)}) \bK_{\textrm{pos}}(\bU_s-\bU_s^{(j)})^H) \geq 0,
\end{equation}
where $ \bK_{\textrm{pos}} = \bK-k_{\min}\bI_{L\NT}\succeq \bzero$, $k_{\min}$  is the smallest eigenvalue of $\bK$, $\bU_s^{(j)}$ satisfies the constraint in \eqref{eq:GeneralQuadProg}, and we drop the superscript ${(n,k)}$ for notational simplicity. Using \eqref{eq:majorInequ}, we obtain
\begin{equation}\label{eq:Minorizer}
  \tr(\bU_s \bK\bU_s^H) \geq 2\Re(\tr(\bU_s^{(j)}\bK_{\textrm{pos}} (\bU_s)^H )) +c_0,
\end{equation}
where $c_0 = 2k_{\min}e_t - \tr(\bU_s^{(j)} \bK (\bU_s^{(j)})^H)$, and we have used the fact that $\tr(\bU_s\bU_s^H) = e_t$. The right-hand side of the inequality is a minorizer of $\tr(\bU_s \bK\bU_s^H)$. Therefore, ignoring the irrelevant constant, the maximization problem of an MM algorithm based on \eqref{eq:Minorizer} is:
\begin{align}\label{eq:GeneralQuadProgMajor}
  \max_{\bU_s}& \ \Re(\tr( \bB^{(j)} \bU_s^H)) \nonumber \\
  \textrm{s.t.}& \ \|\bu_l\|_2^2 = p_s, \quad l=1,\ldots, L\NT,
\end{align}
where $\bB^{(j)} = \bU_s^{(j)}\bK_{\textrm{pos}} $. Note that
\begin{equation}
  \tr( \bB^{(j)} \bU_s^H) = \sum_{l=1}^{L\NT} \bu_l^H \bb^{(j)}_l,
\end{equation}
where $\bb^{(j)}_l$ is the $l$th column of $\bB^{(j)}$.

Hence, the problem in \eqref{eq:GeneralQuadProgMajor} can be split into $L\NT$ independent problems, the $l$th of which is given by
\begin{align}\label{eq:SubProblem}
  \max_{\bu_l}& \ \Re(\bu_l^H \bb^{(j)}_l) \nonumber \\
  \textrm{s.t.}& \ \|\bu_l\|_2^2 = p_s.
\end{align}
We can immediately obtain the optimal solution to \eqref{eq:SubProblem}, which is given by
\begin{equation}
  \bu_l= \sqrt{p_s} \frac{\bb^{(j)}_l}{\|\bb^{(j)}_l\|}.
\end{equation}

Algorithm \ref{Alg:1} summarizes the proposed MM algorithm for solving \eqref{eq:GeneralQuadProg}.
We can use the accelerating scheme in \cite{Varadhan2008SQUAREM} to speed up the convergence of Algorithm \ref{Alg:1}.
\begin{algorithm}[!htp]
  \caption{ \small  Optimization algorithm for \eqref{eq:GeneralQuadProg} }\label{Alg:1}
  \KwIn{$\bK$.}
  \KwOut{$\bU_s$.}
  \textbf{Initialize:} $j = 0$, $\bU_s^{(0)}$, $\bK_{\textrm{pos}} = \bK - k_{\min}\bI_{L\NT}$. \\
    \Repeat{convergence}{
    $\bB^{(j)} = \bU_s^{(j)}\bK_{\textrm{pos}} $\\
    \For{$l=1$ \emph{\KwTo} $L\NT$}
    {$\bu_l = \sqrt{p_s} \frac{\bb^{(j)}_l}{\|\bb^{(j)}_l\|}$}
    $j = j + 1$ \\
    }
\end{algorithm}

\subsection{Randomization Approach}
Once we obtain $\bU_s$ and $\bw$ at the convergence of the proposed algorithm, we can synthesize polyphase waveforms from $\bU_s$ and $\bw$ using a randomization approach. Specifically, let $\bU_s^{\star}$ and $\bw^{\star}$ denote the obtained solution by the proposed algorithm. In addition, denote the number of independent randomizations by $N_r$. Then, at the $n$th randomization, we generate a random vector $\boldsymbol{\chi}_n \sim \mathcal{CN}(\bzero, \bI_{r})$. Let $\widetilde{\boldsymbol{\chi}}_n = (\bU_s^{\star})^H\boldsymbol{\chi}_n$. Clearly,  $\widetilde{\boldsymbol{\chi}}_n \sim \mathcal{CN}(\bzero, (\bU_s^{\star})^H\bU_s^{\star})$. Next we construct a feasible point, denoted by $\check{\bs}_n$, from $\widetilde{\boldsymbol{\chi}}_n$:
\begin{equation}\label{eq:randomConstruction}
  \check{s}_{n}(i) = \sqrt{p_s} \exp(j\Delta\phi\lfloor\arg(\widetilde{{\chi}}_n(i))/\Delta\phi\rceil),
\end{equation}
where $\check{s}_{n}(i)$ and $\widetilde{{\chi}}_n(i)$ represent the $i$th element of $\check{\bs}_n$ and $\widetilde{\boldsymbol{\chi}}_n$, respectively, $i=1,\ldots,L\NT$. 

Next we choose the best polyphase waveform after randomization. One can use a selection method similar to that in \cite{my2016JointDesign}, \ie,
\begin{equation}\label{eq:randomMethod1}
  \bs^{\star} = \arg\max_{\check{\bs}_n}  \frac{\check{\bs}_n^H\bQ_{\txt}(\bw^{\star})\check{\bs}_n} {\check{\bs}_n^H\bR_\txu(\bw^{\star})\check{\bs}_n}.
\end{equation}
However, note that $\bw^{\star}$ is a filter that corresponds to  $\bU_s^{\star}$. Thus, in general it is not optimal for $\check{\bs}_n$. An alternative way to select the polyphase waveforms is based on the cost function of the optimization problem in \eqref{eq:Problem}:
\begin{equation}\label{eq:randomMethod2}
  \bs^{\star} = \arg\max_{\check{\bs}_n}  (\bv_\txt(\check{\bs}_n))^H \bR_\txu^{-1}(\check{\bs}_n)\bv_\txt(\check{\bs}_n).
\end{equation}
Typically, polyphase waveforms with a larger SINR can be obtained by the selection method \eqref{eq:randomMethod2}. However, the computational complexity of this selection method is much larger than that of the selection method \eqref{eq:randomMethod1}, since it involves the calculation of $\bR_\txu(\check{\bs}_n)$ and its inverse.

Algorithm \ref{Alg:2} summarizes the randomization approach for the design of polyphase waveforms. Algorithm \ref{Alg:3} summarizes the overall design method for MIMO radar systems, where $g^{(n)}$  is the objective value of \eqref{eq:JointDesignRepara} at the $n$th (outer) iteration, and $\varepsilon_1$ and $\varepsilon_2$ are predefined small values ($>0$).

\begin{algorithm}[!htp]
  \caption{ \small  Randomization approach for the design of polyphase waveforms}\label{Alg:2}
  \KwIn{$\bU_s^{\star}, \bw^{\star}$.}
  \KwOut{$\bs^{\star}$.}
   \For{$n=1$ \emph{\KwTo} $N_r$}
   {
        Generate $\boldsymbol{\chi}_n \sim \mathcal{CN}(\bzero, \bI_{r})$\\
        $\widetilde{\boldsymbol{\chi}}_n = (\bU_s^{\star})^H\boldsymbol{\chi}_n$\\
        \For{$i=1$ \emph{\KwTo} $L\NT$}
        {
            $\check{s}_{n}(i) = \sqrt{p_s} \exp(j\Delta\phi\lfloor\arg(\widetilde{{\chi}}_n(i))/\Delta\phi\rceil)$\\
        }
   }
   Obtain $\bs^{\star}$ from \eqref{eq:randomMethod1} or \eqref{eq:randomMethod2}.
\end{algorithm}

\begin{algorithm}[!htp]
  \caption{ \small  Polyphase  waveform design algorithm for MIMO STAP}\label{Alg:3}
  \KwIn{$\bV_\txt$, $\{\{\sigma^2_{\textrm{c},p,k},\bV_{\txc,k,p}\}_{k=1}^{N_\txc}\}_{p=-P}^{p=P}$.}
  \KwOut{$\bs^{\star}$.}
  \textbf{Initialize:} $n = 0$, $\bU_s^{(0)}$. \\
  \Repeat{$(g^{(n+1)} - g^{(n)})/g^{(n+1)} \leq \varepsilon_1$}{
  Compute $\bQ_{\txt}(\bU_s^{(n)})$\\
  Compute $\bR_\txu(\bU_s^{(n)})$\\ 
  $\bw^{(n+1)} = \mathcal{P}(\bQ_{\txt}(\bU_s^{(n)}),\bR_\txu(\bU_s^{(n)}))$ \\
  Compute $\bQ_{\txt}(\bw^{(n+1)})$\\ 
  Compute $\bR_\txu(\bw^{(n+1)})$\\ 
  $k=0$\\
  $\bU_s^{(n,k)} = \bU_s^{(n)}$\\
        \Repeat{$(x^{(n,k+1)} - x^{(n,k)})/x^{(n,k+1)} < \varepsilon_2$}{
            $x^{(n,k)} = \frac{\tr(\bU_s^{(n,k)} \bQ_{\txt}(\bw^{(n+1)}) (\bU_s^{(n,k)} )^H)}{\tr(\bU_s^{(n,k)}  \bR_\txu(\bw^{(n+1)})  (\bU_s^{(n,k)} )^H)}$ \\
            $\bK^{(n,k)} = \bQ_{\txt}(\bw^{(n+1)}) - x^{(n,k)}\bR_\txu(\bw^{(n+1)})$\\
            Apply Algorithm \ref{Alg:1} with $\bU_s^{(n,k)}$ as the initial point and let $\bU_s^{(n,k+1)}$ denote the solution at convergence.\\
            $k=k+1$\\
        }
   $\bU_s^{(n+1)} = \bU_s^{(n,k)}$     \\
   $n=n+1$\\
  }
  Use the randomization approach in Algorithm \ref{Alg:2} to obtain $\bs^{\star}$.
\end{algorithm}


\section{Analysis and Discussions} \label{sec:Discussions}
\subsection{Convergence Analysis}
In this subsection, we prove that $\{g^{(n)}\}$ forms a monotonically non-decreasing sequence, \ie, $g^{(n+1)} \geq g^{(n)}$. The proof relies on the ascent property of cyclic optimization, Dinkelbach's transform, and MM. To this end, we can verify that
\begin{equation}
  \tr(\bU_s^{(n,k)} \bK^{(n,k)}(\bU_s^{(n,k)})^H) = 0.
\end{equation}
In addition, we show in Appendix \ref{Apd:B} that the sequence of the objective values during the iterations of Algorithm \ref{Alg:1} is non-decreasing. Thus,
\begin{align}
  &\tr(\bU_s^{(n,k+1)} \bK^{(n,k)}(\bU_s^{(n,k+1)})^H) \nonumber\\
  =&\tr(\bU_s^{(n,k+1)} \bQ_{\txt}(\bw^{(n+1)}) (\bU_s^{(n,k+1)})^H)  \nonumber\\
  &- x^{(n,k)} \tr(\bU_s^{(n,k+1)} \bR_{\txu}(\bw^{(n+1)}) (\bU_s^{(n,k+1)})^H) \geq 0.
\end{align}

Note that
\begin{equation}
  x^{(n,k+1)} = \frac{\tr(\bU_s^{(n,k+1)} \bQ_{\txt}(\bw^{(n+1)}) (\bU_s^{(n,k+1)} )^H)}{\tr(\bU_s^{(n,k+1)}  \bR_\txu(\bw^{(n+1)})  (\bU_s^{(n,k+1)} )^H)}.
\end{equation}
As a result, we have
\begin{equation}\label{eq:xConverge}
  x^{(n,k+1)} \geq x^{(n,k)}.
\end{equation}

Using \eqref{eq:xConverge}, we obtain
\begin{equation}
  g^{(n)} \leq g^{(n+1/2)} = x^{(n,0)}\leq x^{(n,K_{iter})}= g^{(n+1)},
\end{equation}
where $g^{(n+1/2)}$ denotes the objective value associated with $\{\bU_s^{(n)},\bw^{(n+1)}\}$, $K_{iter}$ denotes the number of iterations for the fractional programming problem, and the first inequality holds because of the optimality of $\bw^{(n+1)}$ given $\bU_s^{(n)}$.

Note that the objective of \eqref{eq:JointDesignRepara} is bounded from above. Thus, the objective values provided by the iterates of
Algorithm \ref{Alg:3} are guaranteed to converge to a finite value.
\subsection{On the Computation of $\bR_\txu(\bU_s)$ and $\bR_\txu(\bw)$}
In this subsection, we provide an alternative way of computing $\bR_\txu(\bU_s)$ and $\bR_\txu(\bw)$. First, we introduce the following lemma (for notational simplicity, we discard the dependency of the vectors and matrices on clutter parameters):
\begin{lemma} \label{Lemma:1}
Let $\bx \in \complexC^{L\NT}$. Then
\begin{align}
  (\bd \otimes\bJ_p^T \otimes\bA)\bx
  &= (\bI_M \otimes \bX_p^T \otimes \bI_{N_\txR}) (\bd\otimes\ba_\txT\otimes\ba_\txR), \label{eq:Lemma1_1}\\
  (\bd^H \otimes\bJ_p \otimes\bA^H)\bw &=  (\bJ_p \hat{\bW}\otimes \bI_{N_\txT}) (\bd^*\otimes\ba_\txR^*\otimes\ba^*_\txT),\label{eq:Lemma1_2}
\end{align}
where $\bX_p = \bX \bJ_p$, $\bX \in \complexC^{\NT \times L}$  satisfies $\vec(\bX) = \bx$, $\hat{\bW} \in \mathbb{C}^{L\times M\NR}$  satisfies  $\vec(\hat{\bW}) = (\bI_M \otimes \bK) \bw$, and $\bK \in \mathbb{C}^{L\NR \times L\NR}$ is a commutation matrix\footnote{ An $L\NR \times L\NR$ commutation matrix $\bK$ satisfies $\bK\vec(\bM) = \vec(\bM^T)$ for any $\bM \in \mathbb{C}^{\NR\times L}$. As a result, for $\bv_1 \in \mathbb{C}^{\NR}$ and $\bv_2 \in \mathbb{C}^{L}$, $\bK(\bv_1\otimes \bv_2) = \bv_2 \otimes \bv_1$.}.
\end{lemma}
\begin{IEEEproof}
  See Appendix \ref{Apd:C}.
\end{IEEEproof}

Let $\bc_l$ denotes the $l$th column of $\bU_s^H$, $l=1,\ldots,L\NT$. Then $\bU_s^H\bU_s = \sum_{l=1}^{L\NT} \bc_l\bc_l^H$. Using Lemma \ref{Lemma:1}, one can verify that
\begin{align}
  \bV_{\txc,k,p}\bc_l
  &= (\bd(\omega_{\textrm{c},p,k})\otimes\bJ_p^T \otimes\bA(\theta_{\textrm{c},p,k}))\bc_l \nonumber\\
  &= \widetilde{\bC}_{p,l} \widetilde{\bv}_{\txc,p,k},
\end{align}
where $\widetilde{\bC}_{p,l} = (\bI_M \otimes \bC_{p,l}^T \otimes \bI_{N_\txR})$,  $\bC_{p,l} = \bC_l \bJ_p$, $\bC_l$ is an $\NT \times L$ matrix satisfying $\vec(\bC_l) = \bc_l$, and $\widetilde{\bv}_{\txc,p,k} = \bd(\omega_{\textrm{c},p,k}) \otimes\ba_\txT(\theta_{\textrm{c},p,k})\otimes\ba_\txR(\theta_{\textrm{c},p,k})$. Hence, $\bR_\txu(\bU_s)$ can be rewritten as
\begin{align}\label{eq:RuU2}
  \bR_\txu(\bU_s) =  \sum_{p = -P}^{P}\sum_{l=1}^{r} \widetilde{\bC}_{p,l} \widetilde{\bR}_{\txc,p} \widetilde{\bC}_{p,l}^H  + \bR_{\txJ\txn},
\end{align}
where $\widetilde{\bR}_{\txc,p} = \sum_{k=1}^{N_\txc} \sigma^2_{\txc,p,k}\widetilde{\bv}_{\txc,p,k}\widetilde{\bv}_{\txc,p,k}^H$, which can be computed offline.

By exploiting the sparse structure of $\bV_{\txc,k,p}$ in \eqref{eq:RuU}, the computation of $\bR_\txu(\bU_s)$ has a complexity of $O((2P+1)N_\txc((L\NT)^3+L^2M\NR\NT(\NT+\NR)))$. Compared with \eqref{eq:RuU}, the computation of $\bR_\txu(\bU_s)$ with \eqref{eq:RuU2} has a complexity of $O((2P+1)r(L(M\NR\NT)^2+\NT(LM\NR)^2))$. In some applications, to improve the fidelity of the clutter model, the number of clutter patches $N_c$ is set to a large value. In such situations, calculating $\bR_\txu(\bU_s)$ with \eqref{eq:RuU2} can be more efficient.

Similarly, using \eqref{eq:Lemma1_2}, we can verify that
\begin{equation}
  \bV_{\txc,k,p}^H\bw =\widetilde{\bW}_p   \breve{\bv}_{\txc,k,p}^*,
\end{equation}
where $\widetilde{\bW}_p = \bJ_p \hat{\bW}\otimes \bI_{N_\txT} $, and $\breve{\bv}_{\txc,k,p} = \bd(\omega_{\textrm{c},p,k})\otimes\ba_\txR(\theta_{\textrm{c},p,k})\otimes\ba_\txT(\theta_{\textrm{c},p,k})$. Thus, $\bR_\txu(\bw)$ can be rewritten as
\begin{equation}\label{eq:Ruw2}
  \bR_\txu(\bw) =  \sum_{p = -P}^{P} \widetilde{\bW}_p \breve{\bR}_{\txc,p} \widetilde{\bW}_p^H+ \beta(\bw) \bI_{L\NT},
\end{equation}
where $\breve{\bR}_{\txc,p}=\sum_{k=1}^{N_\textrm{c}} \sigma^2_{\textrm{c},p,k} \breve{\bv}_{\txc,k,p}^*\breve{\bv}_{\txc,k,p}^T$, which also can be computed offline.

The calculation of $\bR_\txu(\bw) $ via \eqref{eq:Ruw} has a complexity of $O((2P+1)N_\txc((L\NT)^2+ LM\NR+L\NT))$. Compared with \eqref{eq:Ruw}, the computation of $\bR_\txu(\bw)$ using \eqref{eq:Ruw2} has a complexity of $O((2P+1)(L(M\NR\NT)^2) + M\NR(L\NT)^2)$. Thus, if $N_\txc$ is large, we can choose \eqref{eq:Ruw2} to reduce the computational complexity.

\subsection{Computational Complexity}

Next we discuss the computational complexity per iteration.
With respect to the optimization of $\bw$ for fixed $\bU_s$, the complexity is mainly determined by the calculation of $\bR_\txu(\bU_s)$ and the eigenvector associated with the largest generalized eigenvalue of  the matrix pair. In the previous subsection, we have analyzed the computational complexity of calculating $\bR_\txu(\bU_s)$, which is dependent on the computation method. For the calculation of the eigenvector associated with the largest eigenvalue of  the matrix pair, the complexity is $O((LM\NR)^3)$. Using the method in \cite{Golub2002generalEig}, the complexity can be reduced to $O((LM\NR)^2)$.
For the optimization of $\bU_s$ given $\bw$, the complexity is proportional to that of Algorithm 1, which is $O(r(L\NT)^2)$. As a comparison, the optimization of $\bR_s$ given $\bw$ with the interior point method has a complexity of $O((L\NT)^{6.5})$  \cite{Ben2001convex}, which is much higher than that of the proposed algorithm. Moreover, interior point methods usually need the formation of a Schur complement matrix, which has a memory requirement of $O((L\NT)^{4})$. Such a significant requirement hinders the application of interior point methods to the design of long codes. 

\subsection{The selection of $r$} \label{Subsec:LowRank}
In this subsection,  we make use of the result in \cite{Huang2010rank} to show that, there exists a low-rank solution to the optimization of $\bR_s$ given $\bw$. To this end, we rewrite the optimization of $\bR_s$ given $\bw$ as follows:
\begin{align}\label{eq:DesignRs}
  \max_{\bR_s}& \    \frac{\tr(\bR_s \bQ_{\txt}(\bw))}{\tr(\bR_s \bR_\txu(\bw))}\nonumber \\
  \textrm{s.t.}& \ \textrm{diag}(\bR_s) = p_s\cdot \bone_{L\NT\times 1}, \quad \bR_s \succeq \bzero.
\end{align}

After applying the Charnes-Cooper transform \cite{charnes1962programming}, we can find the optimal solution by solving the following SDP:
\begin{align}\label{eq:DesignMs}
  \max_{\bM_s,\eta}& \    {\tr(\bM_s \bQ_{\txt}(\bw))}\nonumber \\
  \textrm{s.t.}& \ {\tr(\bM_s \bR_\txu(\bw))} = 1, \eta>0, \nonumber\\
                    &\ \textrm{diag}(\bM_s) = \eta p_s\cdot \bone_{L\NT\times 1}, \quad \bM_s \succeq \bzero.
\end{align}
Denote the optimal solution of \eqref{eq:DesignMs} by $\{\bM_s^{\textrm{opt}}, \eta^{\textrm{opt}}\}$. Then the optimal solution of \eqref{eq:DesignRs} is given by
\begin{equation}
  \bR_s^{\textrm{opt}} = \bM_s^{\textrm{opt}}/\eta^{\textrm{opt}}.
\end{equation}
Define $\widetilde{\bM}_s = \textrm{BlkDiag}(\bM_s;\eta)$, $\widetilde{\bQ}_{\txt}(\bw) = \textrm{BlkDiag}(\bQ_{\txt}(\bw);0) $, $\widetilde{\bR}_\txu(\bw) = \textrm{BlkDiag}(\bR_\txu(\bw);0)$, and $\widetilde{\be}_l=[\be_l;-\sqrt{p_s}]$, $l=1,\ldots,L\NT$, where $\be_l$ is the canonical vector with the $l$th entry equal to 1 and other entries equal to zero. We can verify that \eqref{eq:DesignMs} can be recast as
\begin{align}\label{eq:DesignMsTilde}
  \max_{\widetilde{\bM}_s}& \    {\tr(\widetilde{\bM}_s \widetilde{\bQ}_{\txt}(\bw))}\nonumber \\
  \textrm{s.t.}& \ {\tr(\widetilde{\bM}_s \widetilde{\bR}_\txu(\bw))} = 1, \widetilde{\bM}_s \succeq \bzero, \nonumber\\
                    &\ \tr(\widetilde{\bM}_s \widetilde{\be}_l \widetilde{\be}_l^H)=0, l=1,\ldots,L\NT.
\end{align}

It follows from  \cite[Lemma 3.1]{Huang2010rank} that there is an optimal solution of \eqref{eq:DesignMsTilde} satisfying
\begin{equation}\label{eq:lowRankResult}
  \textrm{rank}(\widetilde{\bM}_s^{\textrm{opt}}) = \textrm{rank}({\bM}_s^{\textrm{opt}})+1 =\textrm{rank}({\bR}_s^{\textrm{opt}})+1 \leq \sqrt{L\NT+1}.
\end{equation}
The above result implies that it is sufficient to consider $r \leq \sqrt{L\NT+1}-1$. This can reduce the computational complexity of the proposed algorithm significantly.


%



\section{Numerical Examples} \label{sec:Examples}
In this section, we provide numerical examples to verify the performance of the proposed algorithm. We consider an airborne MIMO radar system with $\NT=4$ transmit antennas and $\NR=4$ receive antennas. We assume for simplicity that both antenna arrays are ULAs, with inter-element spacing $d_\txT = 2\lambda$ and $d_\txR = \lambda/2$, respectively. In addition, the radar array is steered to broadside. The radar platform is moving at a height of $h_a = 9000$ m and a constant speed of $v_a = 150$ m/s. The transmit waveforms have a carrier frequency of $f_c = 1$ GHz and a bandwidth of $B = 1$ MHz. The total available energy for the waveforms is $e_t=1$ and the code length is $L=13$. The radar transmits $M=16$  pulses in a CPI with a PRF $f_r=1000$ Hz. The target of interest is at a range of $R_t = 12728$ m (which corresponds to an elevation angle $\approx 45^{\circ}$) and an azimuth of $0^{\circ}$ (\ie, the target DOA is $\theta_t=0$). For the clutter model, we consider $P=1$ and $N_\txc = 361$ clutter patches in a range ring. In addition, $\sigma_{\txc,k,p}^2 = 1, p=-P,\ldots,P, k=1,\ldots,N_\txc$. One barrage noise jamming is present with a jammer-to-noise-ratio (JNR) of $35$ dB and a DOA of $30^{\circ}$.  We use \eqref{eq:RuU2} to compute $\bR_
\txu(\bU_s)$ and  \eqref{eq:Ruw2} to compute $\bR_\txu(\bw)$. We assume that the noise is white with $\sigma^2=1$. The number of randomizations in the randomization procedure is $N_r=100$. For the CycSDR algorithm, the involved convex optimization problems are solved by the CVX toolbox \cite{grant2014cvx}.
 Regarding the stopping criterion of the proposed algorithm, we set $\varepsilon_1 = \varepsilon_2 = 10^{-3}$.
Finally, all the analysis is carried out on a standard laptop with Intel Core i7-8550U CPU and 16 GB RAM.

\begin{figure}[!htp]
\centering
\includegraphics[width = 0.4\textwidth]{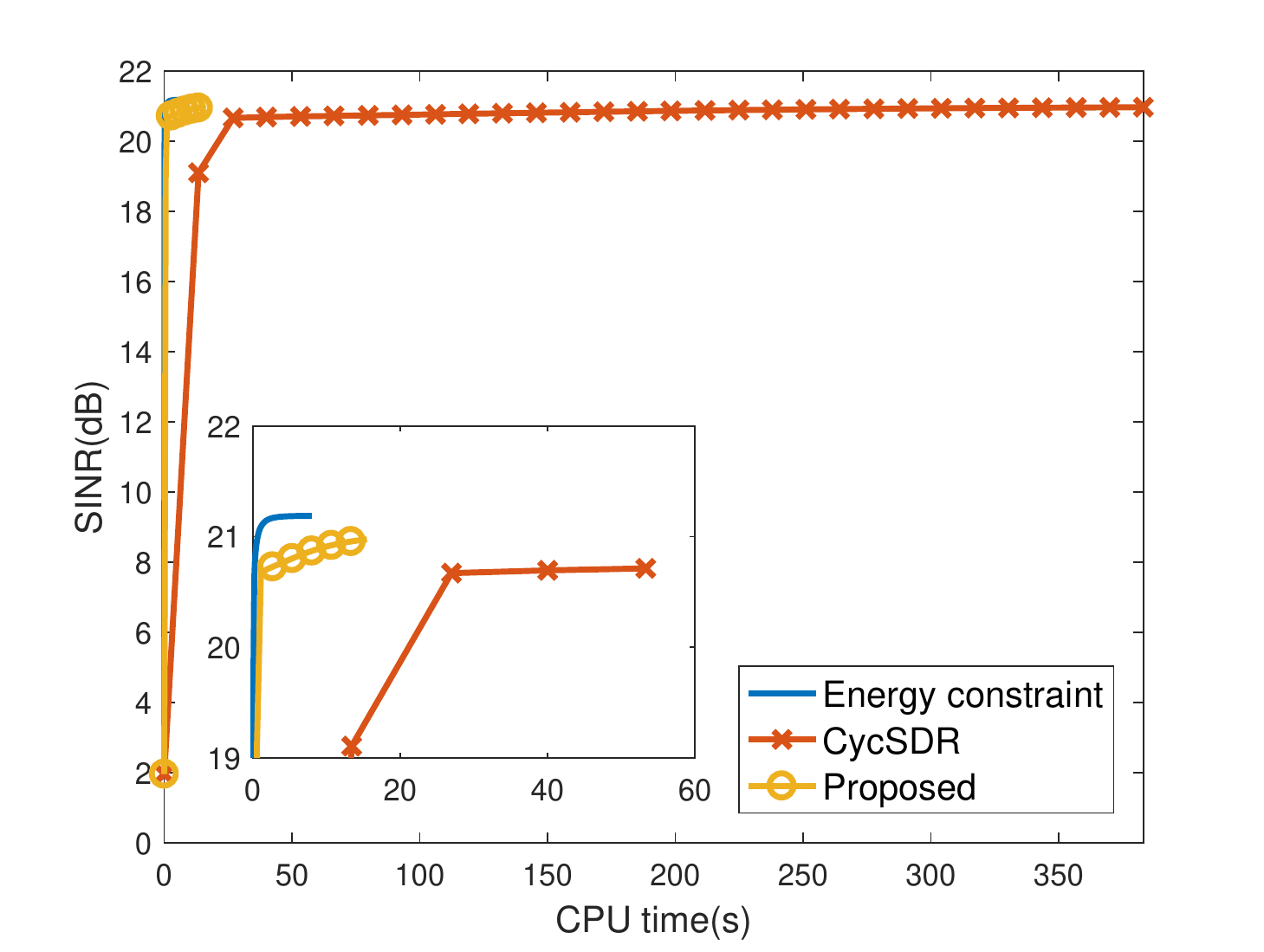}
\caption{Convergence of SINR versus CPU time. $v_t=30$ m/s. $r=6$.}
\label{Fig:1}
\end{figure}

First, we analyze the convergence of the proposed algorithm (before randomization). Fig. \ref{Fig:1} shows the SINR curves for the proposed algorithm and the CycSDR algorithm versus CPU time, where the target velocity is $v_t=30$ m/s (which corresponds to a normalized Doppler frequency of 0.2), and $r= \left\lfloor\sqrt{L\NT+1}\right\rfloor -1 =6$. Herein, we initialize $\bU_s$ by a random matrix with independent and identically distributed (i.i.d.) elements and  the columns normalized according to the constraint, and we initialize $\bR_s$ by   $\bR_s = \bU_s^H\bU_s$. In addition, we plot the SINR associated with the waveforms designed only with the energy constraint (which can be designed by Algorithm 1 in \cite{my2016JointDesign}) as a benchmark. Note the monotonically increasing behavior of the SINR curves and the convergence of both algorithms. The SINRs of the CycSDR algorithm and the proposed algorithm at convergence are $20.98$ dB and $20.97$ dB, which are slightly smaller than that of the energy-constrained  waveforms ($21.19$ dB). Regarding the CPU time needed to reach convergence, the proposed algorithm takes $15.5$ seconds to converge and the CycSDR algorithm takes $383.1$ seconds to converge.  Thus, the proposed algorithm is much faster than the CycSDR algorithm. We believe that the efficiency of the proposed algorithm is due to not only the lower per-iteration computational complexity, but also to the good initial points (from the previous outer iterations, see Algorithm \ref{Alg:3} for details) for the inner iterations.

\begin{figure}[!htp]
\centering
{\subfigure[]{{\includegraphics[width = 0.4\textwidth]{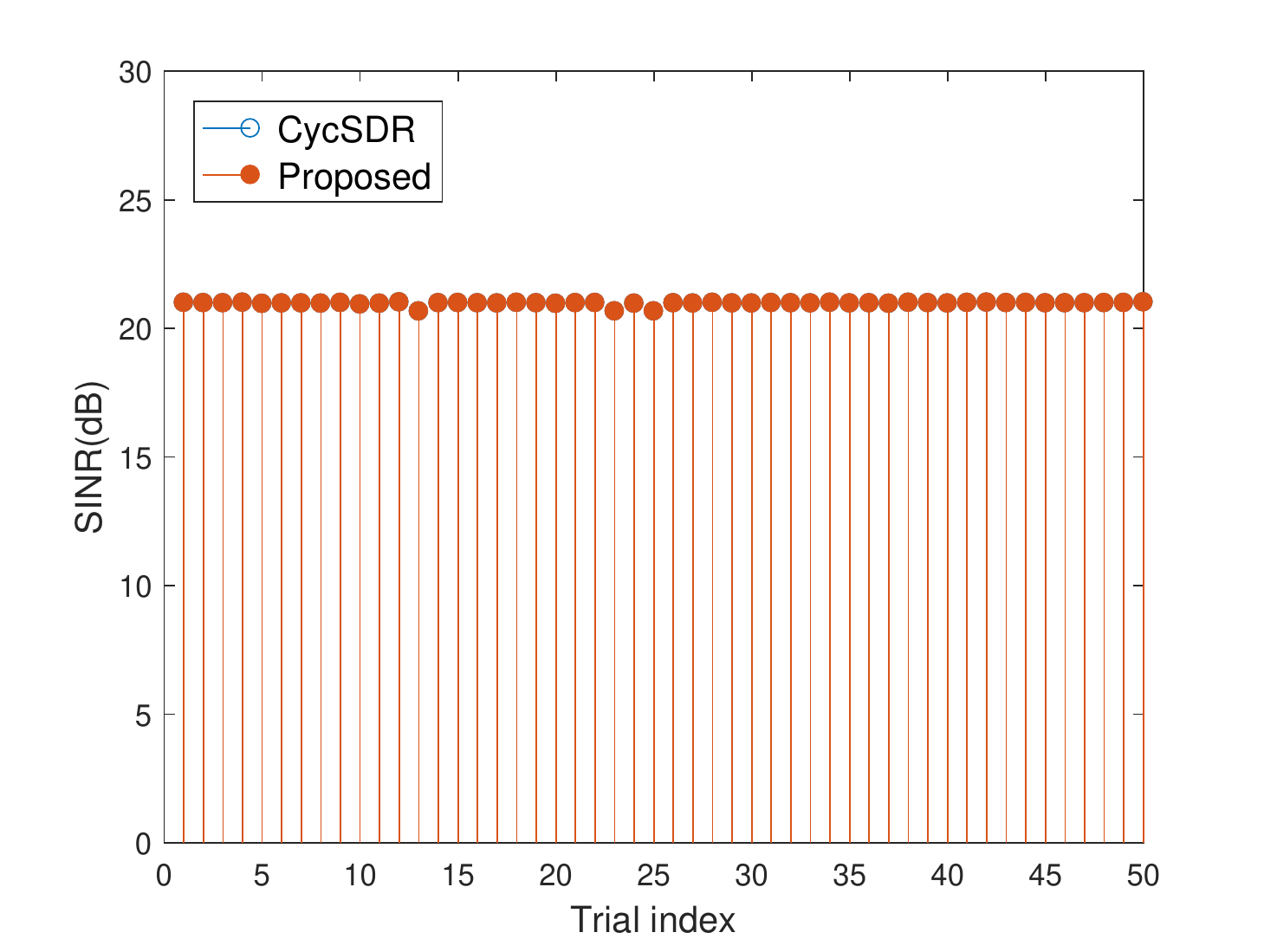}} \label{Fig:2a}} }
{\subfigure[]{{\includegraphics[width = 0.4\textwidth]{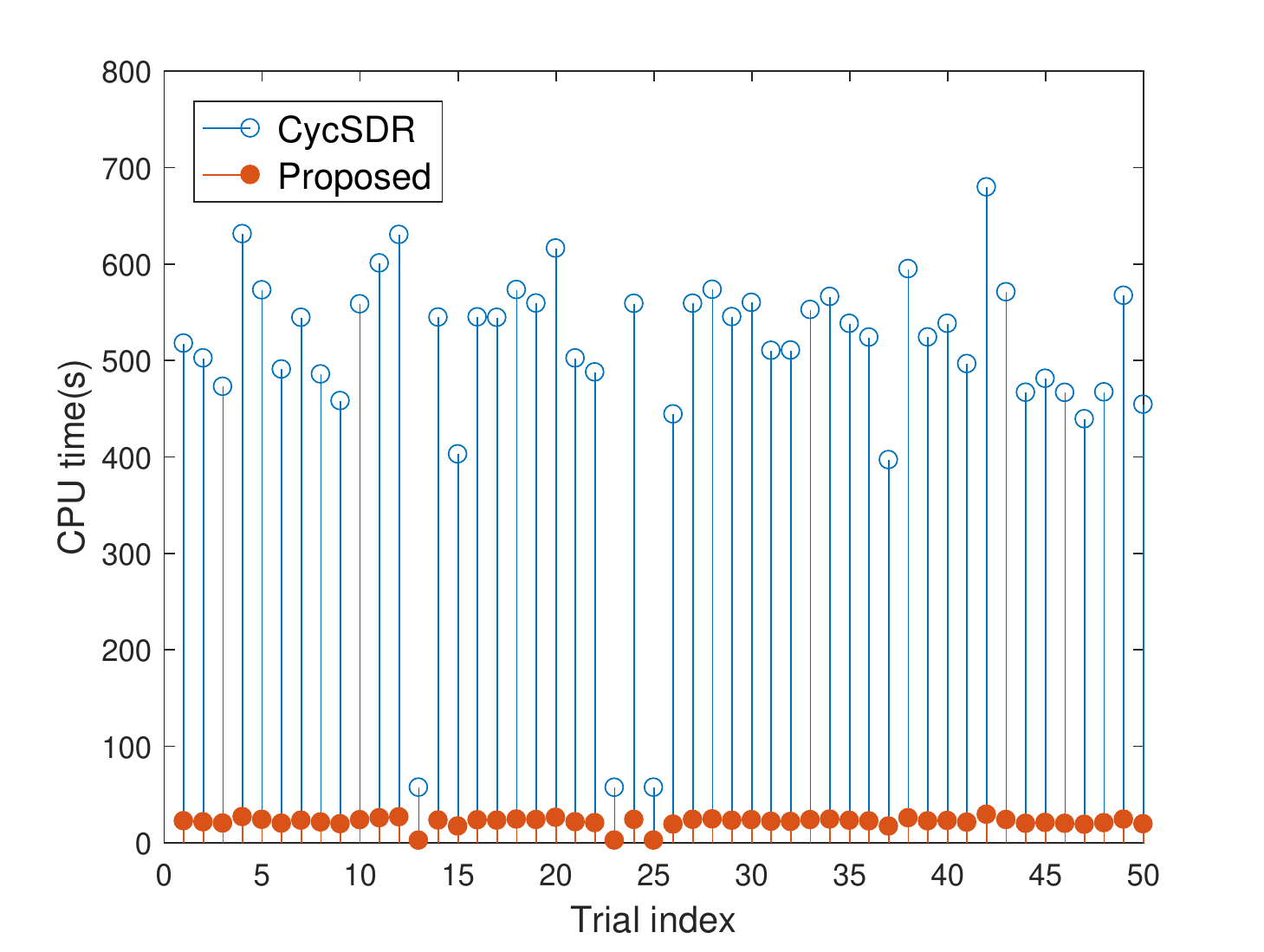}} \label{Fig:2b}} }
\caption{The impact of starting points on SINR and CPU time until convergence. (a) SINR. (b) CPU time. 50 independent Monte Carlo trials. $v_t=30$ m/s. $r=6$.}
\label{Fig:2}
\end{figure}

To further show the efficiency of the proposed algorithm and study the impact of starting points on the performance of the proposed algorithm, we use the same parameters as in Fig. \ref{Fig:1} and conduct $50$ independent Monte Carlo runs using different starting points for $\bU_s$. The associated SINRs and CPU time for the $50$ independent runs are shown in Fig. \ref{Fig:2}. We notice that the SINRs of the proposed algorithm are almost identical to those of the CycSDR algorithm.  In addition, the SINRs of both algorithms are insensitive to the starting points, but the starting point impacts the  convergence speed, and typically, CycSDR is much slower than the proposed algorithm. Since the proposed algorithm attains the same SINR as CycSDR but has a significantly reduced computational complexity, we do not consider CycSDR in the following figures.

\begin{figure}[!htp]
\centering
{\subfigure[]{{\includegraphics[width = 0.4\textwidth]{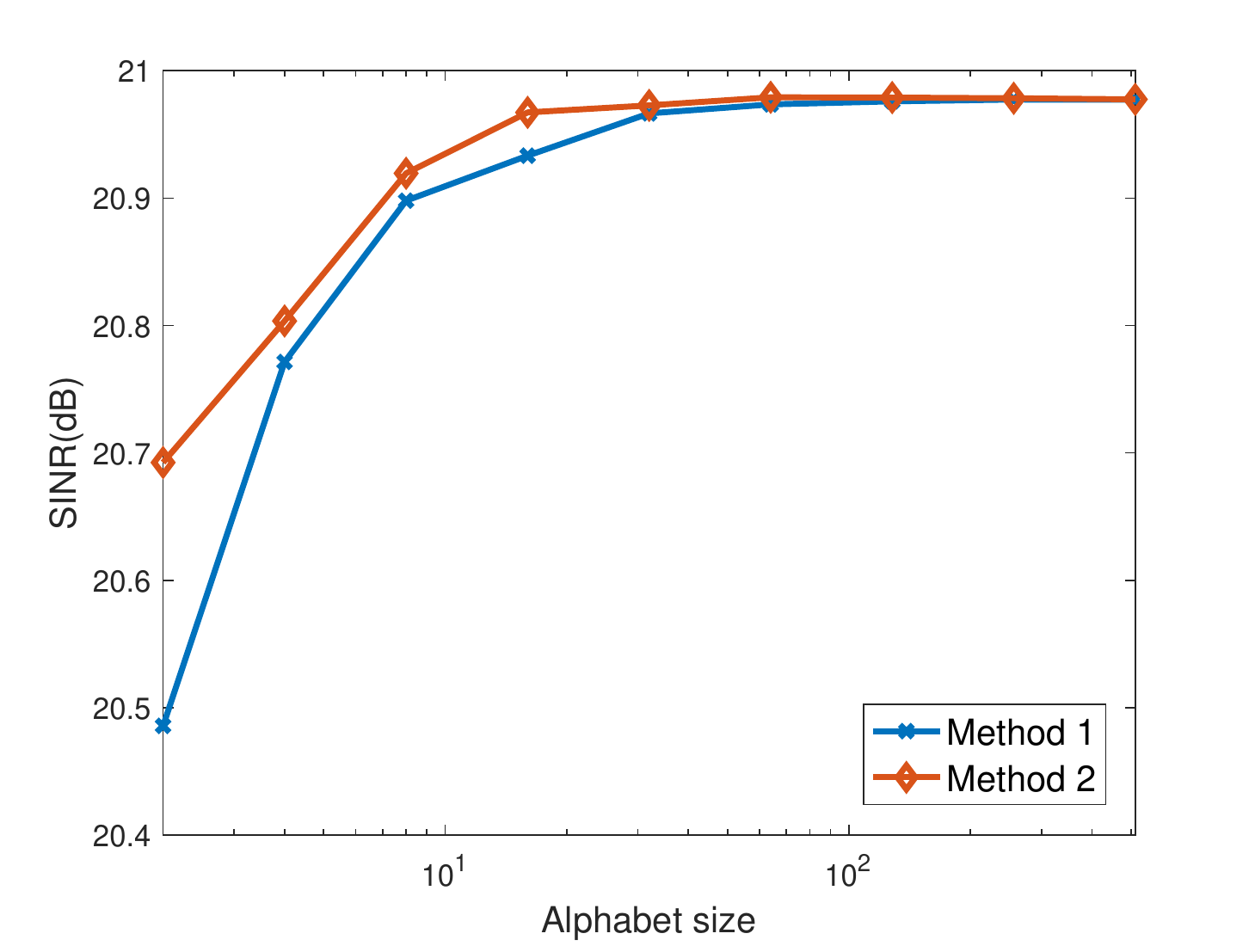}} \label{Fig:3a}} }
{\subfigure[]{{\includegraphics[width = 0.4\textwidth]{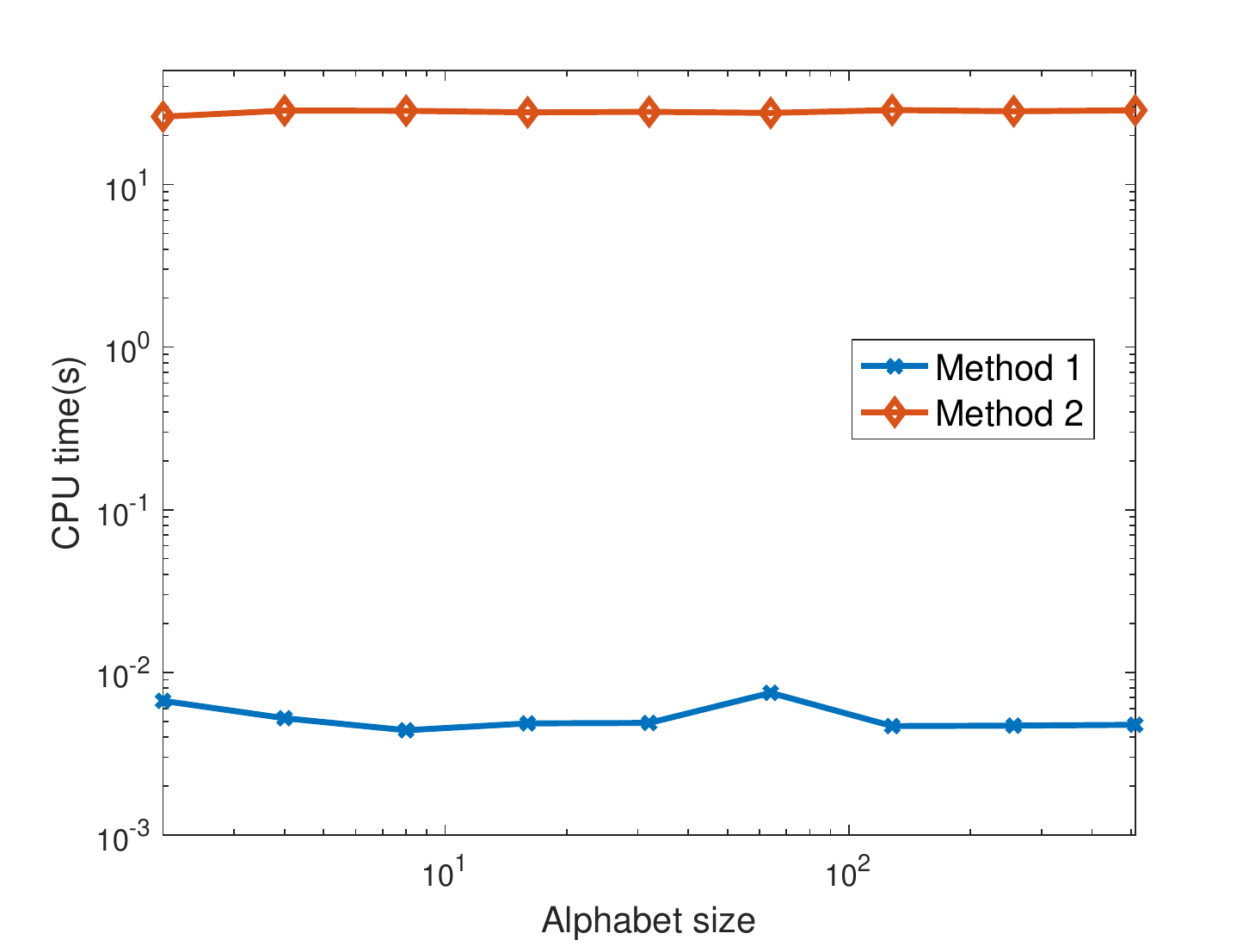}} \label{Fig:3b}} }
\caption{The SINRs and CPU time for different randomization approaches. (a) SINR. (b) CPU time. $v_t=30$ m/s. $r=6$.}
\label{Fig:3}
\end{figure}

Next we analyze the performance of the two randomization approaches proposed in \eqref{eq:randomMethod1}
and \eqref{eq:randomMethod2}.
In Fig. \ref{Fig:3}, we show the performance of the two randomization approaches for $D \in \{2,4,8,16,32,64,128,256,512\}$. We can observe that the waveforms obtained using method $2$ have a larger SINR than those obtained by method 1. However,  the gain is not significant for a large $D$. In addition, the computational complexity of method 2 is several orders of magnitude higher than that of method 1.

\begin{figure}[!htp]
\centering
{\subfigure[]{{\includegraphics[width = 0.4\textwidth]{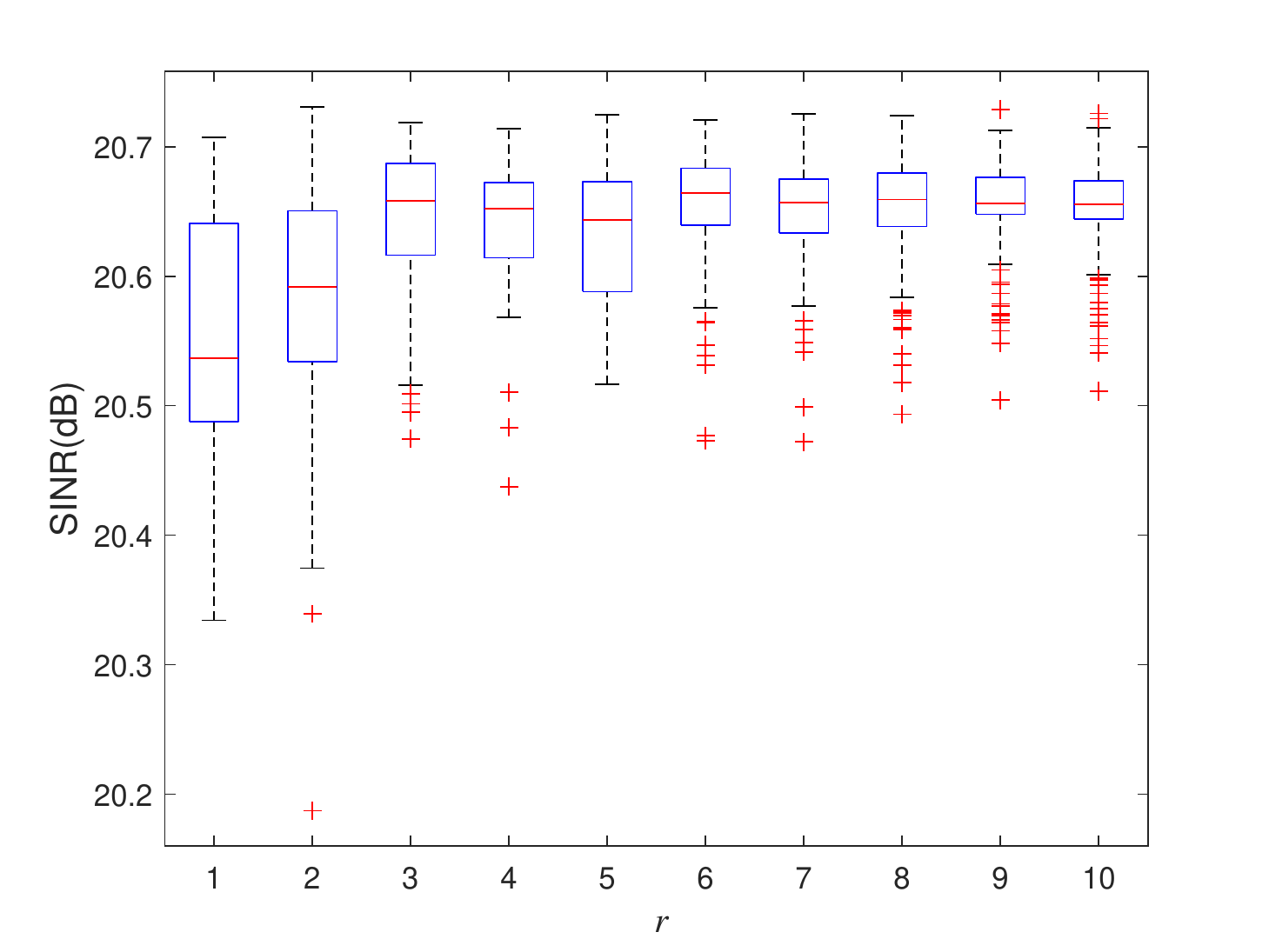}} \label{Fig:4a}} }
{\subfigure[]{{\includegraphics[width = 0.4\textwidth]{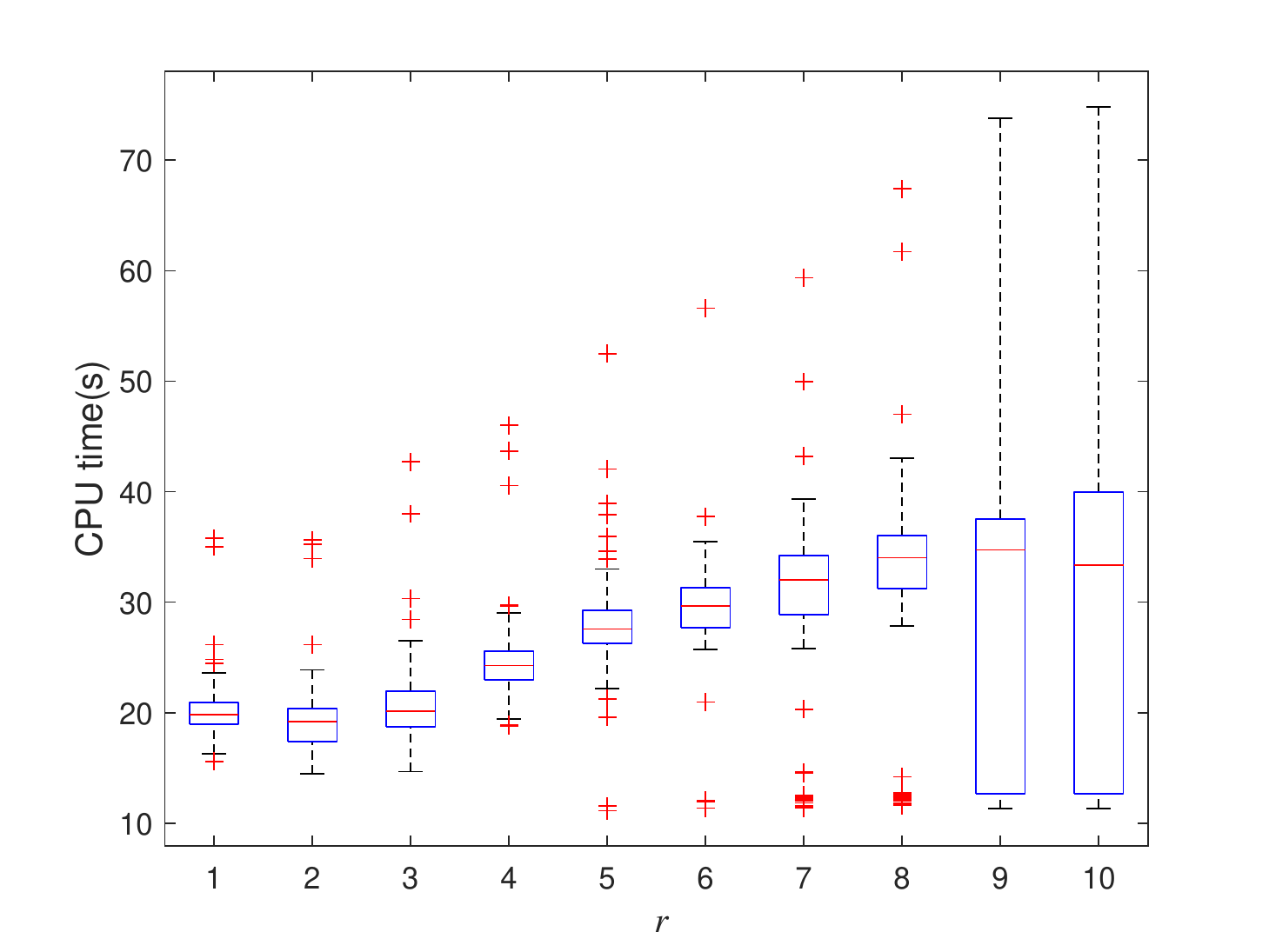}} \label{Fig:4b}} }
\caption{The impact of $r$ on SINR and CPU time. (a) SINR. (b) CPU time. $v_t=30$ m/s. $D=2$.}
\label{Fig:4}
\end{figure}

In the next example, we assess the impact of the rank of $\bU_s$ (\ie, $r$) on the performance of the proposed algorithm. In Fig. \ref{Fig:4}, we plot the SINRs of the designed binary waveforms (\ie, $D=2$) and the associated CPU time (we use method 2 in the randomization process). For each $r$, we generate 100 different $\bU_s$ to initialize the algorithm and the other parameters are the same as those in Fig. \ref{Fig:1}. It can be seen from Fig. \ref{Fig:4} that by increasing $r$, the SINR of the designed binary waveforms is slightly larger but it takes more time to converge. Interestingly, even for $r=1$ we can obtain a satisfactory performance. Indeed, our numerical simulations show that for this typical parameter setup, even if we use a large $r$, the optimized $\bU_s$ during the iterations tends to have a low rank (typically $\leq 3$) and one dominant  singular value. This can partly explain why using $r=1$ can result in a large SINR. However, there is no theoretical guarantee that $r=1$ will work well for more general cases. 

\begin{figure}[!htp]
\centering
{\subfigure[]{{\includegraphics[width = 0.4\textwidth]{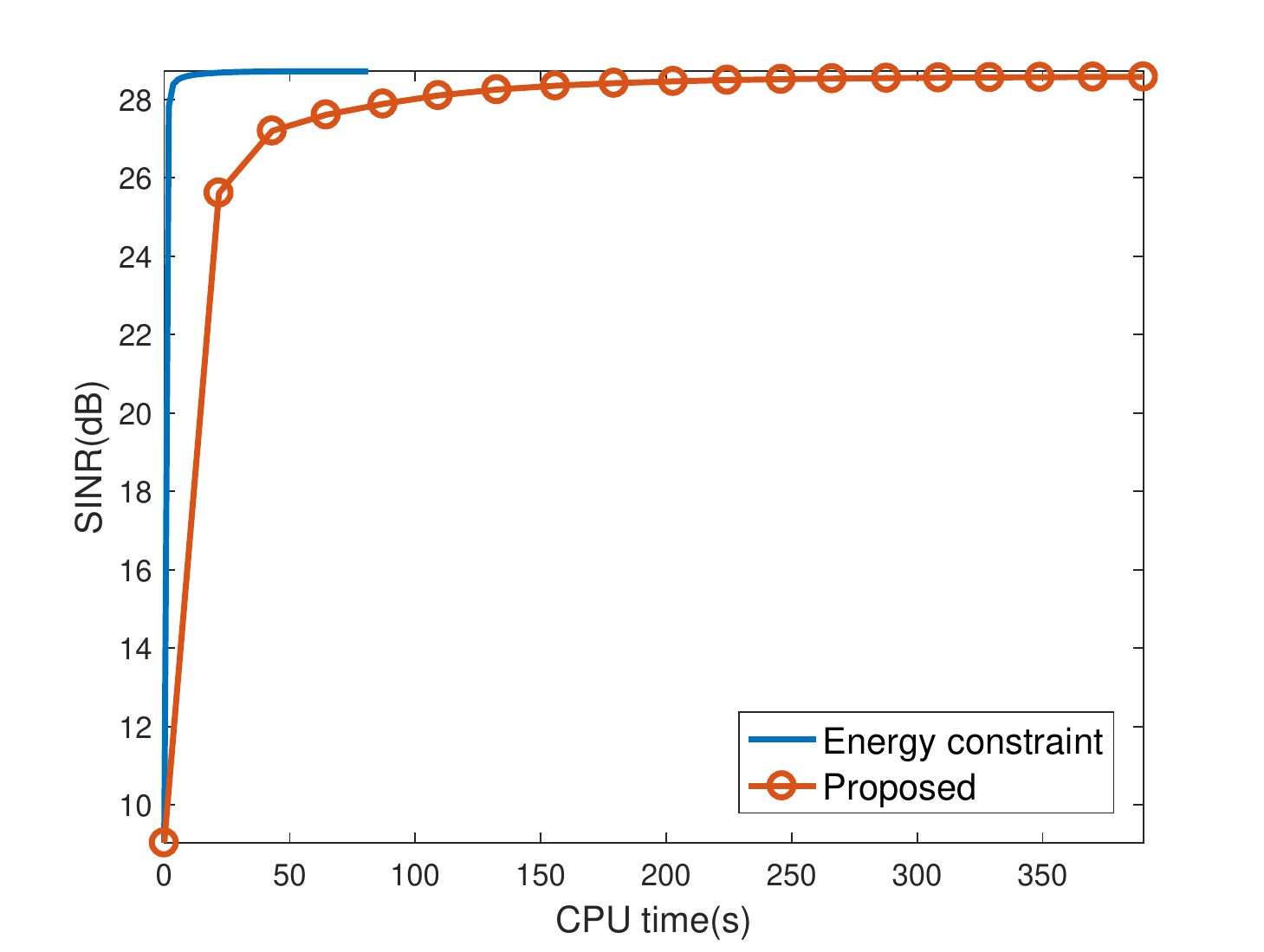}} \label{Fig:5a}} }
{\subfigure[]{{\includegraphics[width = 0.4\textwidth]{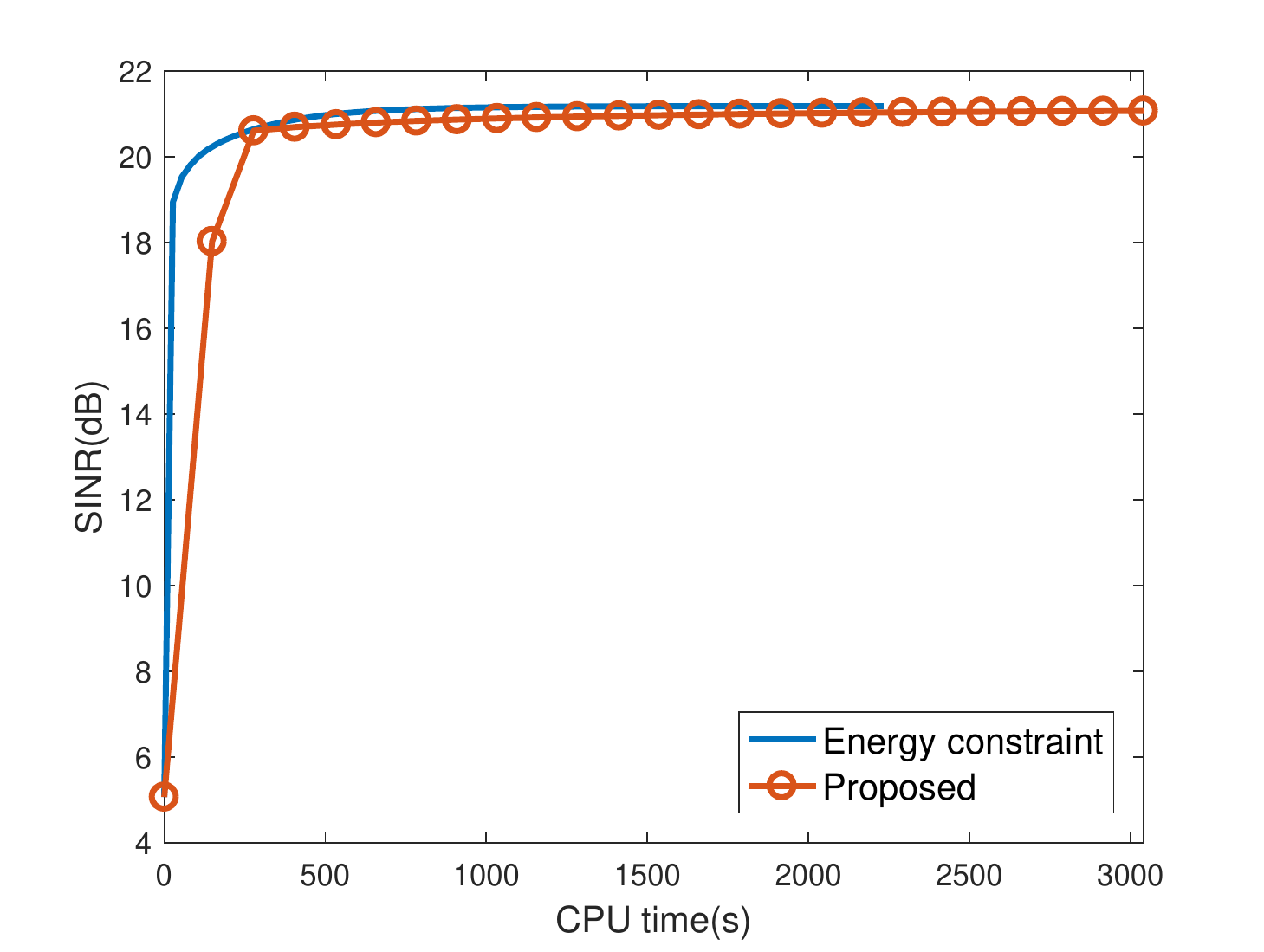}} \label{Fig:5b}} }
\caption{The convergence of SINR versus CPU time for $L\NT = 512$. (a) $\NT=16, L=32$. (b) $\NT=4, L=128$. $v_t=30$ m/s. $D=4$.}
\label{Fig:5}
\end{figure}

To demonstrate that the proposed algorithm can be used to design long waveforms, we let $L\NT=512$ and consider two different cases: $\NT =4, L=128$ and $\NT=16, L=32$. Fig. \ref{Fig:5} shows the convergence of SINR versus CPU time for the proposed algorithm. As a benchmark, the SINR associated with the waveforms designed only with the energy constraint is also plotted. The SINRs for the two designs after randomizations (for $D=4$) are $27.98$ dB and $20.78$ dB, respectively. As expected,  by increasing $L\NT$, the run-time is longer. In addition, we can also notice that increasing the length of the waveforms only results in a slightly larger SINR. Note that by increasing the number of the transmit antennas, we can obtain a higher spatial resolution, which can significantly enhance the SINR at the low Doppler region.

\begin{figure}[!htp]
\centering
{\subfigure[]{{\includegraphics[width = 0.4\textwidth]{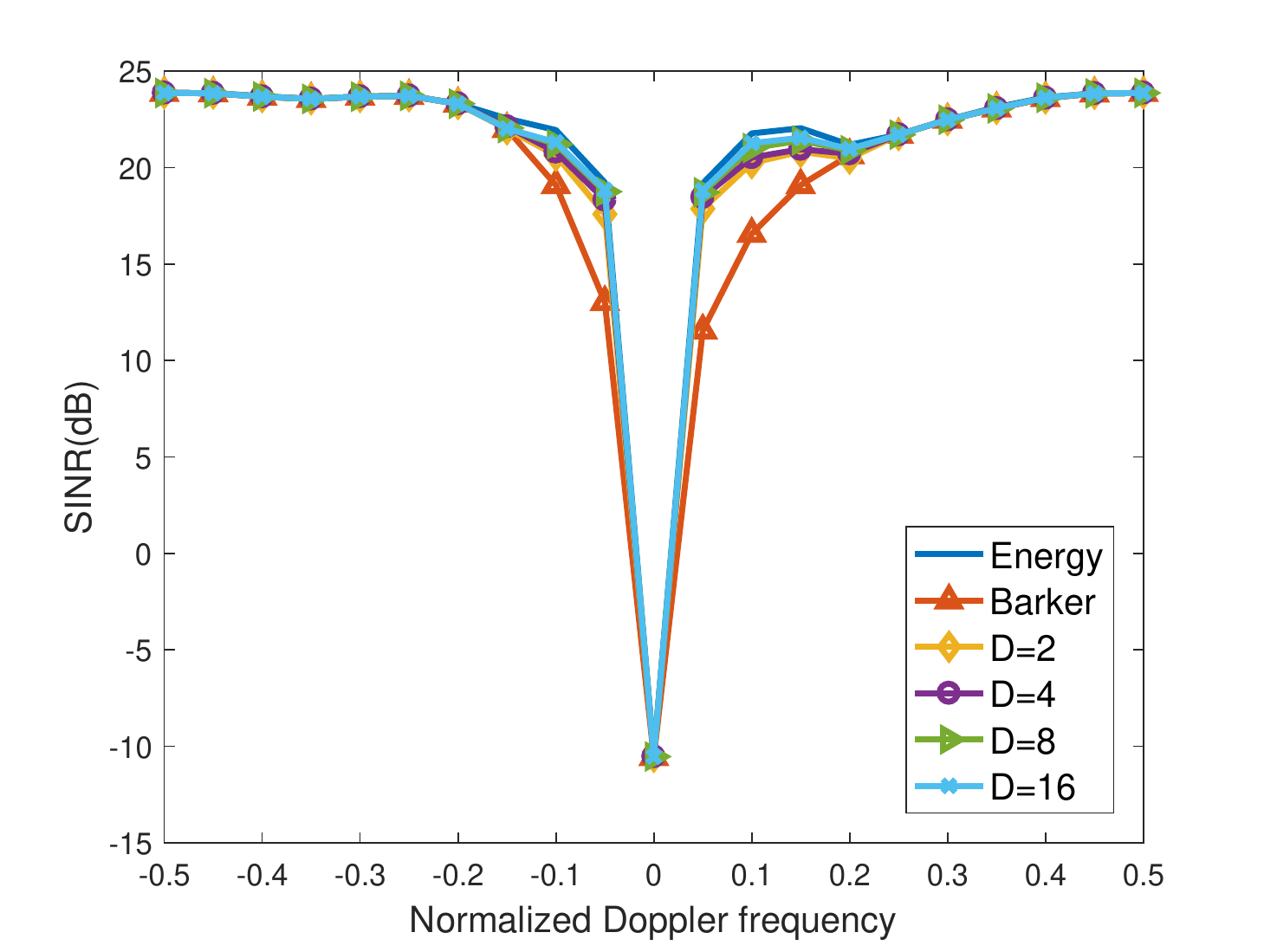}} \label{Fig:6a}} }
{\subfigure[]{{\includegraphics[width = 0.4\textwidth]{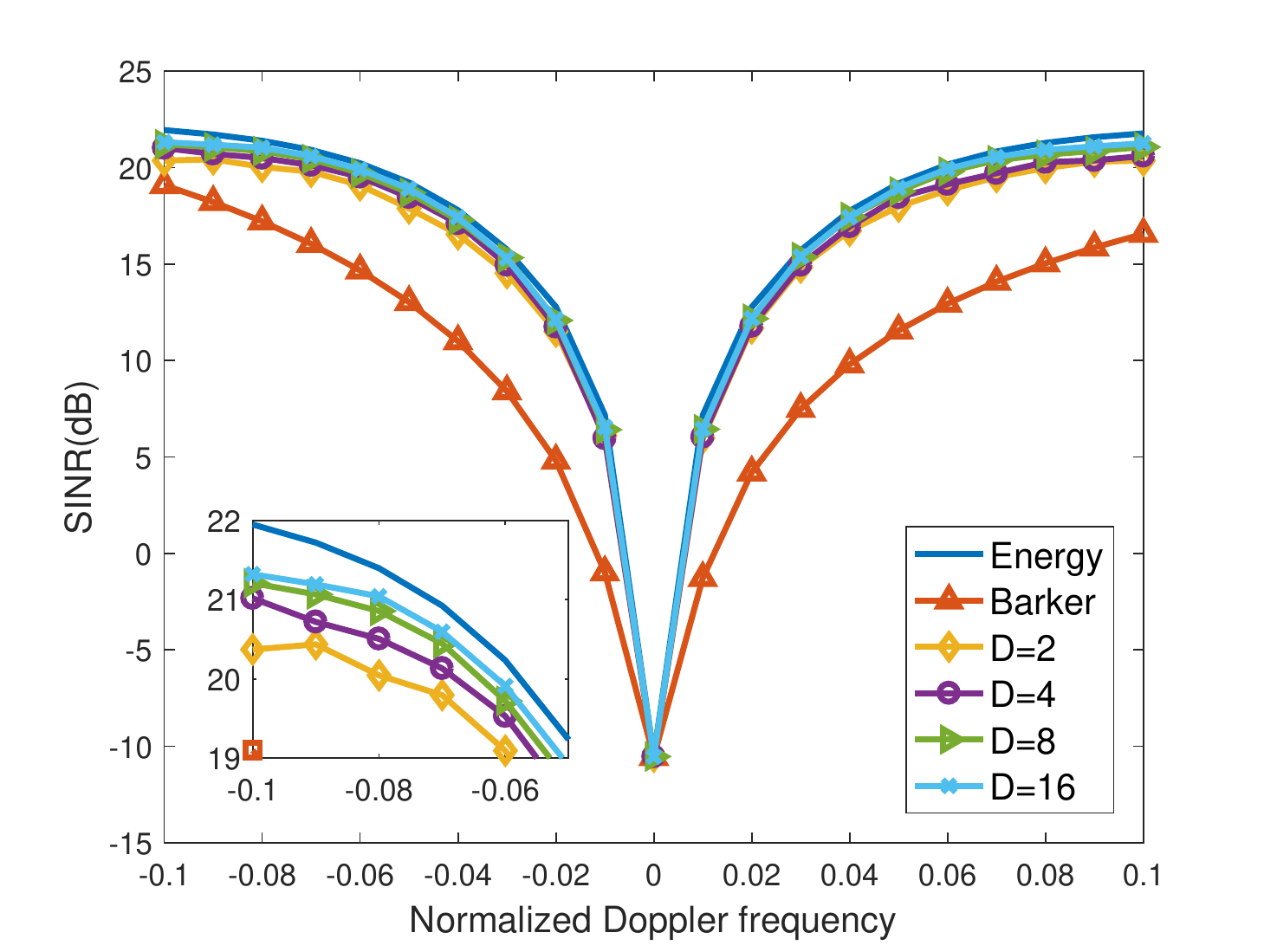}} \label{Fig:6b}} }
\caption{The SINR versus normalized target Doppler frequency. (a) $f_t \in [-0.5, 0.5]$. (b) $f_t \in [-0.1, 0.1]$. $r=6$. }
\label{Fig:6}
\end{figure}

Finally,  we show the SINR of the synthesized polyphase waveforms versus target Doppler frequencies in Fig. \ref{Fig:6a}, where $r=6$, and $D = 2, 4, 8, 16$. Fig. \ref{Fig:6b} shows the corresponding SINR in the low Doppler region. We also plot the SINR of the energy-constrained waveforms and the Barker code for comparison. For the Barker code, the corresponding waveform matrix is $\bS_{\textrm{Barker}} = \ba^{*}(\theta_\txt) \bs_{\textrm{Barker}}^T$, where $\bs_{\textrm{Barker}}$ is the 13-element Barker code. We can observe that the waveforms synthesized by the proposed algorithm is better than the Barker code, especially in the low Doppler region (\eg, the SINR gain using the binary waveforms synthesized by the propose algorithm reaches $7.5$ dB at the normalized Doppler frequency of $0.02$), which is of particular interest for airborne radar. Moreover, increasing $D$ leads to better performance.

%
\section{Conclusion} \label{sec:Conclusion}
We developed an efficient algorithm to design polyphase waveforms for MIMO STAP. The proposed algorithm uses cyclic optimization and relaxation. To reduce the computational complexity and the memory requirement caused by relaxation, we introduced a low-rank reparameterization that was theoretically  justified. Moreover, using Dinkelbach's transform and MM, the reparameterization allows developing an efficient iterative algorithm. We prove that the proposed algorithm has guaranteed convergence of the objective values. Finally, numerical examples showed that, even for the case of synthesizing binary waveforms, the proposed algorithm can achieve a good performance.


\appendices
\section{Proof of \eqref{eq:DesignU}} \label{Apd:A}
First, we can verify that $\bw^H \bV_\txt \bU_s^H\bU_s  \bV_\txt^H \bw= \tr(\bw^H \bV_\txt \bU_s^H\bU_s  \bV_\txt^H \bw) = \tr(\bU_s  \bV_\txt^H \bw\bw^H \bV_\txt \bU_s^H)  = \tr(\bU_s \bQ_{\txt}(\bw) \bU_s^H)$, where we have used the matrix identity that $\tr(\bA\bB) = \tr(\bB\bA)$.
Similarly,
$$\bw^H\bV_{\txc,k,p} \bU_s^H \bU_s \bV_{\txc,k,p}^H \bw= \tr(\bU_s \bV_{\txc,k,p}^H \bw\bw^H\bV_{\txc,k,p} \bU_s^H).$$
In addition,
$ \bw^H \bR_{\txJ\txn}\bw= \bw^H \bR_{\txJ\txn}\bw/e_t \cdot e_t = \beta(\bw)  \tr(\bU_s \bU_s^H)$.
Using the above results, we obtain
\begin{equation}
  \bw^H \bR_\txu(\bU_s)\bw = \tr(\bU_s \bR_\txu(\bw)  \bU_s^H).
\end{equation}
Then the proof of \eqref{eq:DesignU} is completed.

\section{Proof the convergence of Algorithm \ref{Alg:1}} \label{Apd:B}
Let $f^{(j)}$ denotes the objective value of Algorithm \ref{Alg:1} in the $j$th iteration. It is easy to verify that
\begin{subequations}
  \begin{align}
       f^{(j+1)} &= \tr(\bU_s^{(j+1)} \bK(\bU_s^{(j+1)})^H) \\
                &\geq 2\Re(\tr(\bU_s^{(j)}\bK_{\textrm{pos}} (\bU_s^{j+1})^H )) +c_0 \label{eq:ApdB1a}\\
                &\geq 2\Re(\tr(\bU_s^{(j)}\bK_{\textrm{pos}} (\bU_s^{j})^H )) +c_0 \label{eq:ApdB1B} \\
                &= \tr(\bU_s^{(j)} \bK(\bU_s^{(j)})^H) = f^{(j)},
\end{align}
\end{subequations}
where \eqref{eq:ApdB1a} is a consequence of \eqref{eq:Minorizer} and \eqref{eq:ApdB1B} holds because of the optimality of $\bU_s^{(j+1)}$ in the $(j+1)$th iteration. Thus, the sequence of $\{f^{(j)}\}$ is monotonically non-increasing. In addition, it can be verified that $f^{(j)}$ is bounded from above. Thus, the objective values of the iterates of Algorithm \ref{Alg:1} are guaranteed to converge to a finite value.

\section{Proof of Lemma \ref{Lemma:1}} \label{Apd:C}

Note that $(\bd\otimes\bJ_p^T \otimes\bA)\bx = (\bd\otimes\bJ_p^T \otimes\bA) (1\otimes \bx) =  \bd\otimes((\bJ_p^T \otimes\bA)\bx)$.
In addition, $(\bJ_p^T \otimes\bA)\bx = \vec(\bA \bX \bJ_p) = (\bX_p^T \otimes \bI_{N_\txR}) \vec(\bA) = (\bX_p^T \otimes \bI_{N_\txR})(\ba_\txT\otimes \ba_\txR) = (\bX_p^T\ba_\txT)\otimes \ba_\txR$, where $\bX_p = \bX \bJ_p$.
Thus,
\begin{align}
  (\bd\otimes\bJ_p^T \otimes\bA)\bx
  &=  \bd\otimes(\bX_p^T\ba_\txT)\otimes \ba_\txR \nonumber \\
  &= (\bI_M \cdot \bd) \otimes (\bX_p^T\ba_\txT)\otimes (\bI_{N_\txR}\cdot\ba_\txR) \nonumber \\
  &= (\bI_M \otimes \bX_p^T \otimes \bI_{N_\txR}) (\bd\otimes\ba_\txT\otimes\ba_\txR),
\end{align}
which completes the proof of  \eqref{eq:Lemma1_1}.

Define $\bz = (\bd\otimes\bJ_p^T \otimes\bA)\bx$.
Let  $\bz = [\bz_1^T, \bz_2^T ,\ldots, \bz_M^T]^T$ and $\bw = [\bw_1^T, \bw_2^T ,\ldots, \bw_M^T]^T$, then $\bw^H\bz = \sum_{m=1}^M \bw_m^H \bz_m$ and $\bz_m = d_m(\bX_p^T\ba_\txT)\otimes \ba_\txR$, where $d_m$ denotes the $m$th element of $\bd$.
Define $\hat{\bz}_m = d_m\ba_\txR\otimes(\bX_p^T\ba_\txT)$. Then the commutation matrix $\bK \in \mathbb{C}^{L\NR \times L\NR}$ satisfies that $\hat{\bz}_m = \bK \bz_m $.
As a result, $\bw_m^H \bz_m= \bw_m^H \bK^T \bK \bz_m = \hat{\bw}_m^H \hat{\bz}_m$
and $\bw^H\bx$ can be rewritten as $ \bw^H\bz = \sum_{m=1}^M \hat{\bw}_m^H \hat{\bz}_m = \hat{\bw}^H \hat{\bz}$,
where $\hat{\bw}_m = \bK \bw_m$, $\hat{\bw} = [\hat{\bw}_1^T, \hat{\bw}_2^T, \ldots, \hat{\bw}^T_M]^T$, and $\hat{\bz} = [\hat{\bz}_1^T, \hat{\bz}^T_2, \ldots, \hat{\bz}^T_M]^T$. In addition, $\hat{\bz} $ can be written as $ \hat{\bz}  = \bd\otimes\ba_\txR\otimes (\bX_p^T\ba_\txT) = (\bI_{MN_\txR}\otimes \bX_p^T)(\bd\otimes\ba_\txR\otimes\ba_\txT)$.
Note that $ (\bI_{MN_\txR}\otimes \bX_p^*) \hat{\bw} = \vec(\bX_p^* \hat{\bW}) = (\hat{\bW}^T\bJ_p^T \otimes \bI_{N_\txT}) \bx^*$,
where $\hat{\bW} \in \mathbb{C}^{L \times M\NR}$ and $\vec(\hat{\bW} ) = \hat{\bw} $.
Thus, $ \bw^H\bz=\hat{\bw}^H \hat{\bz} = \bx^T (\bJ_p \hat{\bW}^*\otimes \bI_{N_\txT}) (\bd\otimes\ba_\txR\otimes\ba_\txT)$. Since $\bw^H\bz = \bx^T (\bd^T\otimes\bJ_p \otimes\bA^T)\bw^*$, we have
\begin{equation}
  (\bd^H \otimes\bJ_p \otimes\bA^H)\bw =  (\bJ_p \hat{\bW}\otimes \bI_{N_\txT}) (\bd^*\otimes\ba_\txR^*\otimes\ba^*_\txT),
\end{equation}
which completes the proof of \eqref{eq:Lemma1_2}.

%
%

\ifCLASSOPTIONcaptionsoff
  \newpage
\fi

%
%
%
\bibliographystyle{IEEEtran}
\bibliography{IEEEabrv,dicreteCode}

\end{document}